\documentclass[twocolumn]{aastex631}
\usepackage{array}
\usepackage{bm}
\usepackage[slantedGreek]{newtxmath}
\usepackage{subfigure}
\usepackage{graphicx}

\begin{document} 

\title{PKS~2131$-$021 -- Discovery of Strong Coherent Sinusoidal Variations from Radio to Optical Frequencies: Compelling Evidence for a Blazar Supermassive Black Hole Binary}

\author[0000-0001-6314-9177]{S.~Kiehlmann}
\affiliation{Institute of Astrophysics, Foundation for Research and Technology-Hellas, GR-71110 Heraklion, Greece}
\author[0000-0001-5957-1412]{P. V.~de la Parra}
\affiliation{\hbox{CePIA}, Astronomy Department, Universidad de Concepci\'on,  Casilla~\hbox{160-C}, Concepci\'on, Chile}
\author[0000-0002-9545-7286]{A. G. Sullivan}
\affiliation{Kavli Institute for Particle Astrophysics and Cosmology, Department of Physics,
Stanford University, Stanford, CA 94305, USA}
\author[0009-0004-2614-830X]{A. Synani} 
\affiliation{Department of Physics and Institute of Theoretical and Computational Physics, University of Crete, 71003 Heraklion, Greece}
\affiliation{Institute of Astrophysics, Foundation for Research and Technology-Hellas, GR-71110 Heraklion, Greece}
\author[0000-0001-9200-4006]{I. Liodakis}
\affiliation{Institute of Astrophysics, Foundation for Research and Technology-Hellas, GR-71110 Heraklion, Greece}
\affiliation{NASA Marshall Space Flight Center
Huntsville, AL 35808, USA}
\author[0000-0001-7016-1692]{P. Mr{\'o}z}
\affiliation{Astronomical Observatory, University of Warsaw, Al. Ujazdowskie 4, 00-478 Warszawa, Poland}
\author[0000-0002-4478-7111]{S. K.  N{\ae}ss}
\affiliation{Institute of Theoretical Astrophysics, University of Oslo, Norway}
\author[0000-0001-9152-961X]{A.~C.~S.~Readhead}
\affiliation{Institute of Astrophysics, Foundation for Research and Technology-Hellas, GR-71110 Heraklion, Greece}
\affiliation{Owens Valley Radio Observatory, California Institute of Technology,  Pasadena, CA 91125, USA}
\author[0000-0003-0936-8488]{M. C.  Begelman}
\affiliation{JILA, University of Colorado and National Institute of Standards and Technology, 440 UCB, Boulder, CO 80309-0440, USA} 
\affiliation{Department of Astrophysical and Planetary Sciences, 391 UCB, Boulder, CO 80309-0391, USA}
\author[0000-0002-1854-5506]{R. D. Blandford}
\affiliation{Kavli Institute for Particle Astrophysics and Cosmology, Department of Physics,
Stanford University, Stanford, CA 94305, USA}
\affiliation{Cavendish Laboratory, University of Cambridge, 19 JJ Thomson Avenue, Cambridge, CB3 0HE, UK}
\affiliation{Department of Physics and Astronomy, University College London, Gower Street, London WC1E 6BT, UK}
\author[0000-0002-5833-413X]{K. Chatziioannou}
\affiliation{Department of Physics, California Institute of Technology, Pasadena, California 91125, USA}
\affiliation{LIGO Laboratory, California Institute of Technology, Pasadena, California 91125, USA}
\author[0000-0002-5770-2666]{Y. Ding}
\affiliation{Cahill Center for Astronomy and Astrophysics, California Institute of Technology, Pasadena, CA 91125, USA}
\author[0000-0002-3168-0139]{M. J. Graham}
\affiliation{Division of Physics, Mathematics, and Astronomy, California Institute of Technology, Pasadena, CA 91125, USA}
\author[0000-0002-4226-8959]{F. Harrison}
\affiliation{Cahill Center for Astronomy and Astrophysics, California Institute of Technology, Pasadena, CA 91125, USA}
\author{D. C. Homan}
\affiliation{Department of Physics and Astronomy, Denison University, Granville, OH 43023, USA}
\author[0000-0002-2024-8199]{T. Hovatta}
\affiliation{Finnish Centre for Astronomy with ESO (FINCA), University of Turku, FI-20014 University of Turku, Finland}
\affiliation{Aalto University Mets\"ahovi Radio Observatory,  Mets\"ahovintie 114, 02540 Kylm\"al\"a, Finland}
\author[0000-0001-5390-8563]{S. R. Kulkarni}
\affiliation{Cahill Center for Astronomy and Astrophysics, California Institute of Technology, Pasadena, CA 91125, USA}
\author[0000-0003-1315-3412]{M. L. Lister}
\affiliation{Department of Physics and Astronomy, Purdue University, 525 Northwestern Avenue, West Lafayette, IN 47907, USA}
\author[0000-0002-4985-3819]{R. Maiolino}
\affiliation{Kavli Institute for  Cosmology, University of Cambridge, Madingley Road, Cambridge, CB3 0HA, UK}
\author[0000-0002-5491-5244]{W. Max-Moerbeck} 
\affiliation{Departamento de Astronomía, Universidad de Chile, Camino El Observatorio 1515, Las Condes, Santiago, Chile}
\author[0009-0000-9963-6874]{B. Molina}
\affiliation{\hbox{CePIA}, Astronomy Department, Universidad de Concepci\'on,  Casilla~\hbox{160-C}, Concepci\'on, Chile}
\author[0000-0001-6421-054X]{C. P. O'Dea}
\affiliation{Department of Physics and Astronomy, University of Manitoba, Winnipeg, MB R3T 2N2,  Canada}
\affiliation{Center for Space Plasma \& Aeronomic Research, 
University of Alabama in Huntsville,
Huntsville, AL 35899  USA}
\author[0000-0002-0870-1368]{V. Pavlidou} 
\affiliation{Department of Physics and Institute of Theoretical and Computational Physics, University of Crete, 71003 Heraklion, Greece}
\affiliation{Institute of Astrophysics, Foundation for Research and Technology-Hellas, GR-71110 Heraklion, Greece}
\author[0000-0001-5213-6231]{T. J. Pearson}
\affiliation{Owens Valley Radio Observatory, California Institute of Technology,  Pasadena, CA 91125, USA}
\author[0000-0003-2483-2103]{M. F. Aller}
\affiliation{Department of Astronomy, University of Michigan, 323 West Hall, 1085 S. University Avenue, Ann Arbor, MI 48109, USA}
\author[0000-0002-5983-6481]{C. R. Lawrence}
\affiliation{Jet Propulsion Laboratory, California Institute of Technology, 4800 Oak Grove Drive, Pasadena, CA 91109, USA}
\author{T.~J.~W.~Lazio}
\affiliation{Jet Propulsion Laboratory, California Institute of Technology, 4800 Oak Grove Drive, Pasadena, CA 91109, USA}
\author[0000-0003-2454-6828]{S. O'Neill}
\affiliation{Department of Physics, Princeton University, Jadwin Hall, Princeton,
08540, NJ, USA.}
\author[0000-0002-8850-3627]{T. A. Prince}
\affiliation{Division of Physics, Mathematics, and Astronomy, California Institute of Technology, Pasadena, CA 91125, USA}
\author[0000-0002-7252-5485]{V. Ravi}
\affiliation{Owens Valley Radio Observatory, California Institute of Technology,  Pasadena, CA 91125, USA} 
\author[0000-0001-5704-271X]{R. A. Reeves}
\affiliation{\hbox{CePIA}, Astronomy Department, Universidad de Concepci\'on,  Casilla~\hbox{160-C}, Concepci\'on, Chile}
\author[0000-0002-8831-2038]{K. Tassis} 
\affiliation{Department of Physics and Institute of Theoretical and Computational Physics, University of Crete, 71003 Heraklion, Greece}
\affiliation{Institute of Astrophysics, Foundation for Research and Technology-Hellas, GR-71110 Heraklion, Greece}
\author[0000-0002-4162-0033]{M. Vallisneri}
\affiliation{Jet Propulsion Laboratory, California Institute of Technology, 4800 Oak Grove Drive, Pasadena, CA 91109, USA}
\author[0000-0001-7470-3321]{J. A. Zensus}
\affiliation{Max-Planck-Institut f\"ur Radioastronomie, Auf dem H\"ugel 69, D-53121 Bonn, Germany}

\begin{abstract}
Haystack and Owens Valley Radio Observatory (OVRO) observations recently revealed strong, intermittent, sinusoidal total flux-density variations that maintained coherence between  1975 and 2021 in the blazar PKS~2131$-$021 ($z=1.283)$. This was interpreted as possible evidence of  a supermassive black hole binary (SMBHB). Extended observations through~2023 show coherence over 47.9~years, with an observed period $P_\textrm{15 GHz}=(1739.8  \pm 17.4) \, {\rm days}$. We reject, with $p$-value = $2.09 \times 10^{-7}$, the hypothesis that the variations are due to random fluctuations in the red noise tail of the power spectral density. There is  clearly a physical phenomenon in PKS~2131$-$021 producing coherent sinusoidal flux density variations. We find the coherent sinusoidal intensity variations extend from below 2.7~GHz to optical frequencies, from which we derive an observed period  $P_\textrm{optical}=(1764  \pm 36)$ days.  Across this broad frequency range there is a smoothly-varying monotonic phase shift in the sinusoidal variations with frequency.  Hints of  periodic variations are also observed at $\gamma$-ray energies.  The importance of well-vetted SMBHB candidates  to searches for gravitational waves is pointed out. We estimate  the fraction of blazars that are SMBHB candidates to be  $>1$ in 100. Thus monitoring programs covering tens of thousands of blazars  could  discover hundreds of SMBHB candidates.
\end{abstract}

  

\section{INTRODUCTION}

The 40~m Telescope of the Owens Valley Radio Observatory (OVRO) has been dedicated to monitoring the total 15\,GHz flux densities of $\sim 1830$ blazars since 2008 \citep{Richards2011} on a 3--4~day cadence. The first statistically robust report of a
 strong supermassive black hole binary (SMBHB) candidate (PKS~2131$-$021) showing sinusoidal variations in its radio light curve was presented by \citet[hereafter Paper~1]{2022ApJ...926L..35O}.

 SMBHBs are a direct result of hierarchical structure formation and an important ingredient for our understanding of the evolution of the Universe \cite[e.g.,][]{Volonteri2003,Maiolino2023,2024MNRAS.531..355U}.  In spite of the  significant effort that has been put into finding SMBHBs \cite[see][for a recent review]{Dorazio2023}, the number of compelling candidates remains small. This is particularly true for blazars, active galactic nuclei (AGN) with jets oriented toward the observer on Earth \citep{Blandford2019,Hovatta2019}. Until recently, the most compelling blazar SMBHB  candidate with a separation $\ll 1$ pc had been OJ~287, identified through more than 100~years of optical observations showing an outburst every $\sim$11 years due to the secondary black hole crossing the accretion disk of the primary \citep{Sillanpa1988,2016ApJ...819L..37V,2021MNRAS.503.4400D}. All-sky surveys in optical and $\gamma$-rays  have produced more candidates (e.g., PG~1302$-$102, \citealp{Graham2015}; PG~1553+113, \citealp{Ackermann2015,2024ApJ...976..203A,2024MNRAS.529.3894M}), and future facilities like the Vera C.~Rubin observatory are expected to produce many more \cite[e.g.,][]{Xin2021}. 
 
 In the radio regime, periodicity searches have been limited because of the scarcity of facilities capable of regularly observing a large number of blazars. The OVRO 40-m Telescope blazar program \citep{Richards2011} is currently the largest radio flux-density monitoring program. Taking advantage of the quality and length of the OVRO light curves, we have undertaken a  search  for statistically significant periodic and quasi-periodic oscillations (QPOs). In Paper~1 we presented the first results on PKS~2131$-$021, demonstrating highly statistically significant periodicity in this blazar. These results showed that the observed periodicity    is unlikely to be due to random fluctuations and the red noise tail in the variability spectrum. In view of the stability of its period, PKS~2131$-$021 is a strong SMBHB candidate. In this paper, we present further evidence, which we find compelling, supporting the hypothesis of the SMBHB nature of PKS~2131$-$021.  

We analyze the radio to optical light curves of PKS~2131$-$021 in terms of a  model due to Blandford (see Paper~1), which ascribes the sinusoidal variations in PKS~2131$-$021  to aberration caused by the orbital motion in an SMBHB of the black hole producing the jet. This same model had been developed by \citet{2017MNRAS.465..161S} in the context of a ``spine-sheath'' model, but we missed the underlying similarity. We will refer to this as the ``Kinematic Orbital'' model (KO model).

As discussed in Paper~1, blazars showing evidence of periodicities  should not be considered bona fide periodic or QPO candidates unless it can be shown that there is a very low \textit{global} probability ($p$-value $<1.3 \times 10^{-3}$), taking into account the look-elsewhere effect, that the periodicity is due to red noise tail fluctuations, through a careful analysis of the power spectral density (PSD) and through simulations that reproduce both the observed PSD and the observed probability density function (PDF) seen in the light curve. Such analysis is essential for assessing the chance occurrence probability of sinusoidal  features.

    The factors supporting the hypothesis that the blazar PKS~2131$-$021 is an \hbox{SMBHB}, based on 15~GHz light curves from the Haystack and Owens Valley Radio Observatories up to 2021, were presented in Paper~1.  In this paper, we present new 15~GHz data that confirm predictions based on observations obtained up to 2021, and show that there is no need to ``adjust'' the model: the new OVRO data match the model predictions \textit{without altering its original parameters}, thereby fulfilling a classic test of a scientific model.

We also analyze the multi-frequency light curves of PKS~2131$-$021 using Haystack observations from 1975--1983 \citep{1986AJ.....92.1262O} at~2.7~GHz, 7.9~GHz, 15.5~GHz and 31.4~GHz; 15~GHz observations from \hbox{OVRO}; and \hbox{ALMA} observations at 91.5~GHz, 104~GHz, and 345~GHz.  In addition, we analyze infrared data from WISE, optical data from the Zwicky Transient Facility (ZTF), X-ray data from Swift-XRT, and $\gamma$-ray data from the Fermi Gamma-ray Space Telescope, and show that sinusoidal variations, coherent with the radio variations, are clearly seen at optical wavelengths, and, furthermore,  hints of coherent periodic variations are seen at both infrared wavelengths and  $\gamma$-ray energies.

\begin{figure*}[!t]
   \centering
   \includegraphics[width=1.0\linewidth]{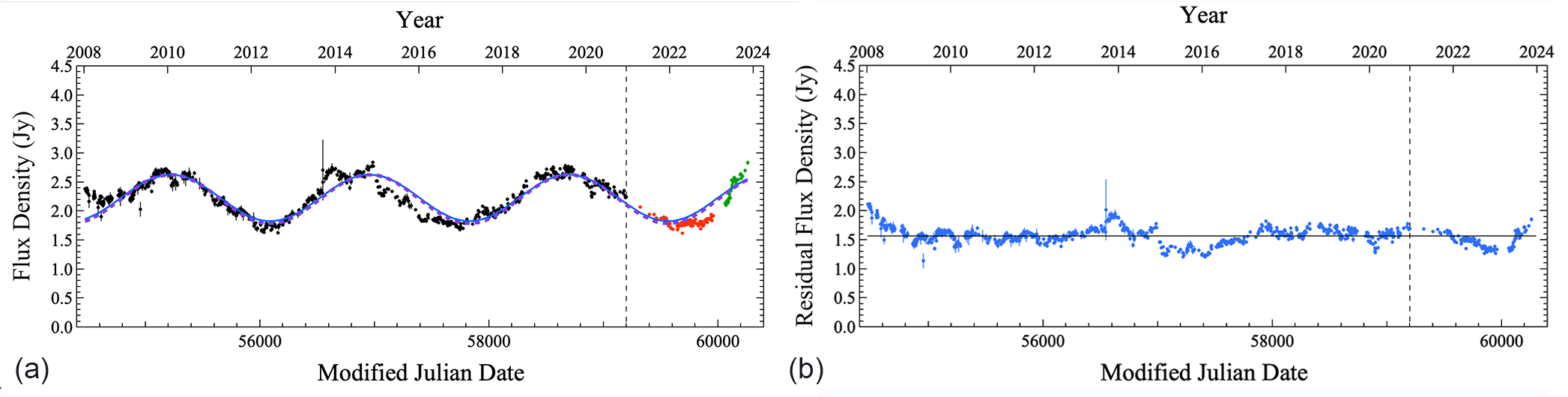}
      \caption{Verified prediction in the OVRO 15 GHz light curve of PKS~2131$-$021. (a)~Black points: The OVRO 15~GHz light curve  presented in Paper~1, which extends up to the vertical dashed line; blue curve: the least-squares sine-wave fit to the Haystack+OVRO data given in Table 1 of  Paper~1, extrapolated to the end of~2023. Note the divergence of the data (red points) from the fit. The prediction at this point was that the light curve would revert to the sine wave. The more recent 2023 observations (green points) show that the observed curve has returned to the extrapolated fit (blue curve), as predicted in Paper~1. The purple dashed sine curve, which is almost indistinguishable from the blue sine curve, is the best  least squares fit, given in Table \ref{tab:bestfit}, to the Haystack data plus all of the OVRO data to the end of 2023. (b) The residual 15~GHz light curve after subtraction of the sinusoidal component (purple dashed sine wave) plotted in~(a). The black horizontal line indicates the mean flux density in the residual.}
         \label{plt:lightcurves}
\end{figure*}

A critical aspect of the sinusoidal variations in PKS~2131$-$021 reported in Paper 1 is the fact that they disappeared completely for 19 years (1984--2003). This raises two questions: (i) given this disappearance, how can we be sure that there is an underlying ``clock'' in PKS~2131$-$021, and (ii) what is responsible for the disappearance and  re-appearance of the sinusoidal variations?  The answer to (i) is that when the sinusoidal variations returned after the 19-year gap they had both the same period and  were ``in step'' with the original sine wave.  This indicates that the ``clock'' had continued ticking throughout.  The rigorous statistical analysis of Paper 1 showed that the hypothesis that the two sets of sinusoidal variations (from 1975 to 1983 and from 2003 to 2021) arose from random fluctuations with a  red noise power spectral density (PSD) could be rejected, with a  $p$-value$=1.58 \times 10^{-6}$ or a significance level of  $4.66 \sigma $. One possible  answer to (ii), which was proposed in Paper 1, is that the sporadic nature of the observed sinusoidal variations could be due to variations in the fueling supplied to the region of the jet inside the compact core component that is producing the sinusoidal variations.

This paper is organized as follows: in Section~\ref{sec:continuation} we discuss the continuation of the 15 GHz sinusoidal variations in PKS~2131$-$021 for another 2 yr; in Section~\ref{sec:correlated1}  we discuss the effects of correlated noise on our analysis; in Section~\ref{sec:statistics} we discuss the statistical significance of our results; in Section~\ref{sec:specpol} we  discuss the radio spectrum, structure,  and polarization  to provide background for the later sections; in Section~\ref{sec:multifrequency} we present evidence that strong coherent sinusoidal variations are seen from 2.7 GHz to optical frequencies, with a smoothly-varying monotonic phase shift as a function of frequency; in Section~\ref{sec:x-ray} we discuss the X-ray observations  and the prospects for future X-ray observations;  in Section~\ref{sec:gamma} we discuss the $\gamma$-ray emission; in Section~\ref{sec:model} we discuss the multifrequency observations in terms of the  KO model; in Section~\ref{sec:fraction} we discuss the fraction of blazars that are SMBHB candidates; in Section~\ref{sec:discussion} we summarize our findings and discuss their implications. There are four appendices:  in Appendix \ref{sec:sineoptical} we discuss the determination of the parameters of the sinusoidal optical light curves; in Appendix \ref{sec:correlated} we discuss the effects of non-white noise deviations from the sinusoidal features  in the light curves; in Appendix \ref{sec:brightness} we discuss the Doppler factor for the compact radio component;  in Appendix \ref{sec:higher2} we discuss the implications of the absence of higher harmonics in the light curve.

For consistency with our other papers, we assume the following cosmological parameters: $H_0 = 71$\, km\,s$^{-1}$\, Mpc$^{-1}$, $\Omega_{\rm m} = 0.27$, $\Omega_\Lambda = 0.73$  \citep{2009ApJS..180..330K}.  None of the conclusions would be changed were we to adopt the best model of the Planck Collaboration  \citep{2020AandA...641A...6P}.

\section{The Continuation of the 15 GHz Sine Wave Signal in PKS~2131$-$021}\label{sec:continuation}

The OVRO 15~GHz light curve of PKS~2131$-$021  is shown in Fig.~\ref{plt:lightcurves}, where the vertical dashed line marks the extent of our calibrated data (up to MJD~59197) analyzed in Paper~1.

In Fig.~\ref{plt:lightcurves}(a) the blue curve shows a least squares sine wave fit (see section~\ref{sec:period-fit}) to the data prior to MJD 59197 (black points) extrapolated forward to the end of 2023 (MJD 60310). As can be seen, by mid-January 2023 (MJD 59959)  the  new data appeared at first to be diverging from the extrapolated sine curve  (red points), but they returned to it in 2023 (green points), thus fulfilling the prediction, from Paper~1, of a constant long-term period. The observed 47.9 year continuity, when combined with the Haystack data, is unprecedented in the light curves of blazars.  In Fig.~\ref{plt:lightcurves}(b) we show the residual variations in PKS~2131$-$021 after subtraction of the sinusoidal component.

It is important to note that the significant departures from the sinusoidal variation in Fig.~\ref{plt:lightcurves}(a) in 2014, 2016 and 2022 do not undermine the SMBHB interpretation of this object.  These departures are most clearly shown in the  light curve with the sine wave subtracted in Fig.~\ref{plt:lightcurves}(b). There is nothing unusual in the blazar light curve of  Fig.~\ref{plt:lightcurves}(b) \citep[see e.g.][]{1972MmRAS..77..109M}.  It is  mildly varying, and we have many such examples in  the $\sim$1830 blazars in the 40 m Telescope monitoring sample.  Thus it is to be \textit{expected} that blazar SMBHBs will show deviations from pure sinusoidal variations due to the non-sinusoidally varying components, some of which are seen clearly as coming from different components in the very long baseline interferometry  (VLBI) maps.  This point is addressed in detail in Paper~1.

\begin{figure}[!tb]
   \centering
   \includegraphics[width=0.85\linewidth]{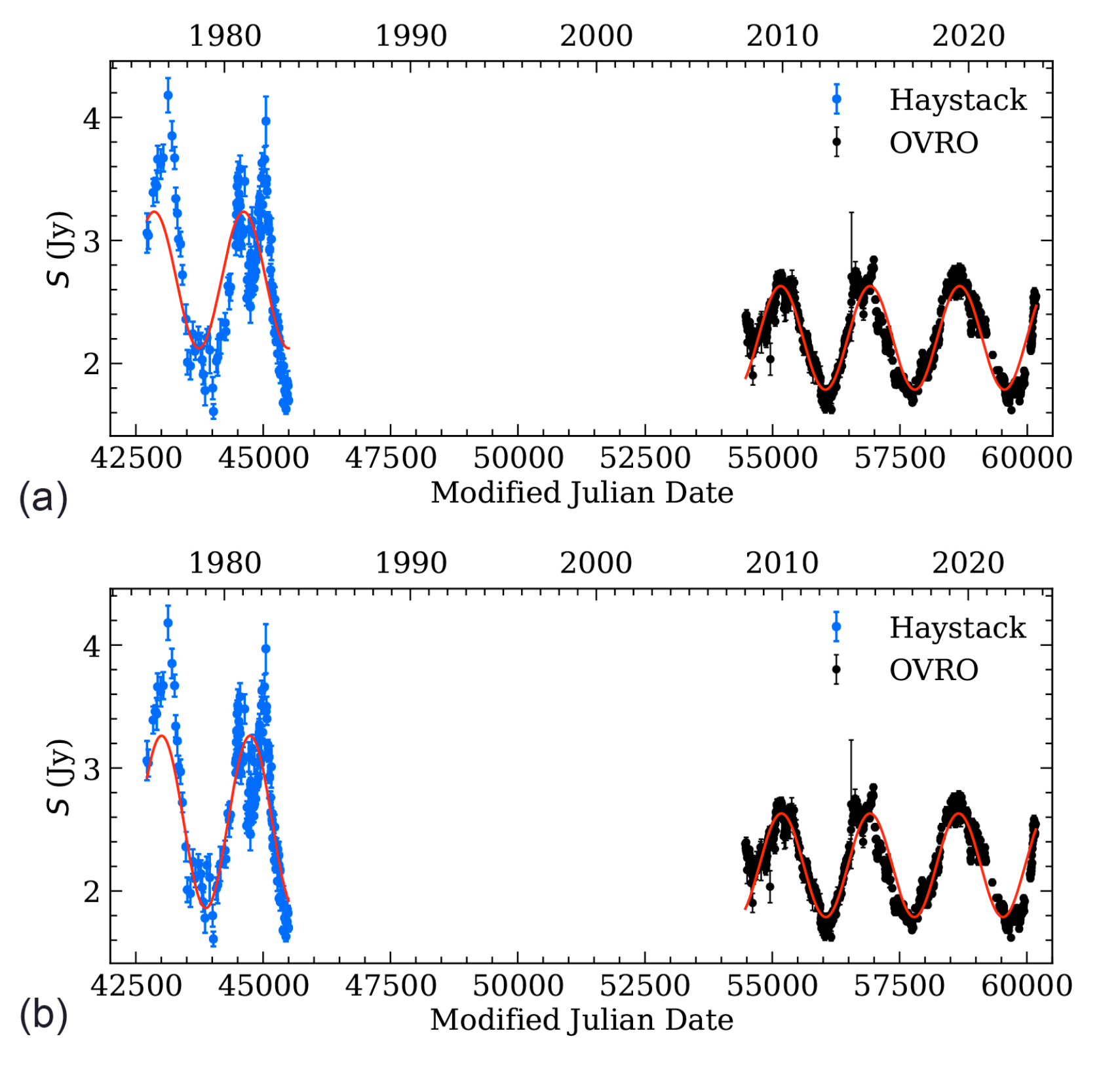}
      \caption{Sine wave fits (red curves) to the combined Haystack observations from~1975 to~1983 \citep{1986AJ.....92.1262O} and OVRO observations from~2008 to the end of 2023.  The amplitudes have been permitted to vary between the two data sets, but the period is derived from the combined data set. (a)~Fit to the unweighted data. (b)~Fit to the weighted data (see text). }
         \label{plt:lightcurves2}
\end{figure}

\begin{table}[!tb]
\centering
\caption{The Least Squares OVRO Sine Wave}
\begin{tabular}{lr}
\hline \hline
Parameter& Value\\
\hline
$P$ (days)         & $1739.2 \pm 1.2$    \\
$\phi_0$ (rad)     & $-0.084 \pm 0.016$  \\
$A$ (Jy)           & $0.4195 \pm 0.0066$ \\
$S_0$ (Jy)         & $2.2084 \pm 0.0046$ \\
$\sigma_0$ (Jy)    & $0.1227 \pm 0.0034$ \\
\end{tabular}
\tablecomments{The fitted parameters are explained in Sec.~\ref{sec:period-fit}. The period is that of the weighted Haystack+OVRO data (see text) shown by the red sine wave in Fig.~\ref{plt:lightcurves2}(b). This period is used as a prior in obtaining the parameters above for the least squares sine wave fit to the OVRO data shown by the purple dashed curve in Fig.~\ref{plt:lightcurves}(a).}
\label{tab:bestfit}
\end{table}

\subsection{The OVRO+Haystack period fit}
\label{sec:period-fit}
We fitted the following functional model to the data:
\begin{equation}
S_{{\rm model}, i} = A \sin(\phi_i-\phi_0) + S_0 + n_i,
\end{equation}
where $\phi_i = 2\pi(t_i-t_0)/P$ is the phase of the $i$th data point, $\phi_0$ is the phase at reference time $t_0$, $A$ is the amplitude, $S_0$ is the flux density offset, and $n_i$ is the noise contribution to the $i$th data point. We allow separate values for $A$ and $S_0$ for the OVRO and Haystack sections of the data.
As discussed in Paper~1, the cause of the difference in the amplitudes of the sine waves in the Haystack data and the OVRO data is unknown, but, given that we ascribe the disappearance of the sinusoidal fluctuations from 1984 to 2003 to a drop in fueling, we suggest that the different amplitudes in these two epochs could be due to different levels of fueling.

The noise term $n_i$ includes observational noise and source intrinsic variability that is not part of the sinusoidal variations. The non-sinusoidal components are effectively a type of correlated noise as far as the fit is concerned, and this correlation should ideally be taken into account in the fit. In the following we apply two noise models. One treats the noise terms as uncorrelated, the other models the correlation. We will show that the fitted parameters are consistent, but the corresponding uncertainties depend on the choice of the noise model.

\subsection{White Noise Model Fit}

First, we treated all noise terms as independent Gaussian noise, $n_i \sim \mathbb{N}(0, \sigma_i^2 + \sigma_0^2)$, where $\sigma_i$ are the data uncertainties and $\sigma_0$ is the source intrinsic noise level, which may be different for OVRO and Haystack fits. The combination of the Haystack and OVRO 15\,GHz data is shown in Fig.~\ref{plt:lightcurves2}, together with two sine wave fits. Between these two epochs the sinusoidal variation disappeared (see Paper~1). In Fig.~\ref{plt:lightcurves2}~(a) we show a sine-wave fit in which each data point is given equal weight. While the fit is quite good, it is clear that there is a slight offset in phase between the sine wave and the Haystack data. The Haystack data are particularly relevant with respect to any phase offsets because of the long time interval of 25 years that separates them from the OVRO data. For the purpose of determining the underlying periodicity, and hence the degree of coherence in the combined data set, in Fig.~\ref{plt:lightcurves2}~(b) we have assigned a weight of 850/154 = 5.5 to each of the Haystack observations, corresponding to the number of data points in the Haystack data (154) and the OVRO data (850). The weights of the OVRO data points are all 1. We refer to this fit as a ``weighted fit''. This has brought the sine wave into alignment with the Haystack data, with no discernible difference in the fit to the OVRO data. The corresponding period is $P=(1739.2\pm1.2)$ days, which may be compared with the period of $P=(1737.9\pm2.6)$ days obtained in Paper~1. The period for the weighted fit in the $z=1.283$ rest frame of the host galaxy is $P_0\!=\!P/(1+z)\!=\!761.8\pm 0.5$ days.

Table~\ref{tab:bestfit} shows the parameters of the best fit to the OVRO data alone, where we fixed the period at $P=(1739.2\pm1.2)$ days, as estimated above from the weighted OVRO+Haystack fit.

\section{The Effects of Correlated Noise}\label{sec:correlated1}

In Fig. \ref{plt:lightcurves3} we show the light curves of PKS~2131$-$021 from 2.7 GHz to optical frequencies. In this section we discuss the light curves from centimeter to sub-millimeter wavelengths shown in Figs. \ref{plt:lightcurves3}(a) and (b). We discuss the optical light curves of \ref{plt:lightcurves3}(c) in \S \ref{sec:optical}.
In Fig.\ref{plt:lightcurves3}(a) we show the Haystack multifrequency light curves from \citet{1986AJ.....92.1262O}. We downloaded ALMA data from the ALMA calibrator website. These are shown in Fig. \ref{plt:lightcurves3}(b). The color-coded curves are least-squares fits of sine waves to the data, derived by the procedure presented in Paper~1 and in Appendix~\ref{sec:correlated}.

We now consider the effects of correlated noise in these multi-frequency light curves. Our adopted noise model is based on light curve simulations that are further explained in Sec.~\ref{sec:statistics}. Details of the model fit are given in Appendix~\ref{sec:correlated}. With this model, we  get a maximum-posterior period of $P = (1739.8 \pm 17.4)$ days. As explained in Appendix~\ref{sec:correlated} the uncertainty is likely to be overestimated.
Note that this estimate has almost exactly the same central value as the weighted white noise fit of $P = (1739.2\pm 1.2)$ of the previous section, showing that the period we measure is robust, and demonstrating that, when the correlation of the non-sinusoidal variations is taken into account, no higher weighting of the Haystack data is required in order to bring the phases into alignment across the 19-year gap in the sinusoidal variations. 

Note that the white noise model fit estimates  error bars  that are lower by about an order of magnitude than those of the correlated noise fit. An analysis carried out by Hincks et al. (in preparation) on sub-millimeter
blazar light curves, shows that the uncertainties provided by the sine
wave fitting approach are reasonably accurate in the presence of Type-A
variations, but better knowledge is needed of SMBHB
phenomenology to determine the uncertainties to better than a factor of two or
three.

\subsection{The Correlated Noise Phase Shift with Frequency}
\begin{figure*}[!ht]
   \centering
   \includegraphics[width=1.0\linewidth]{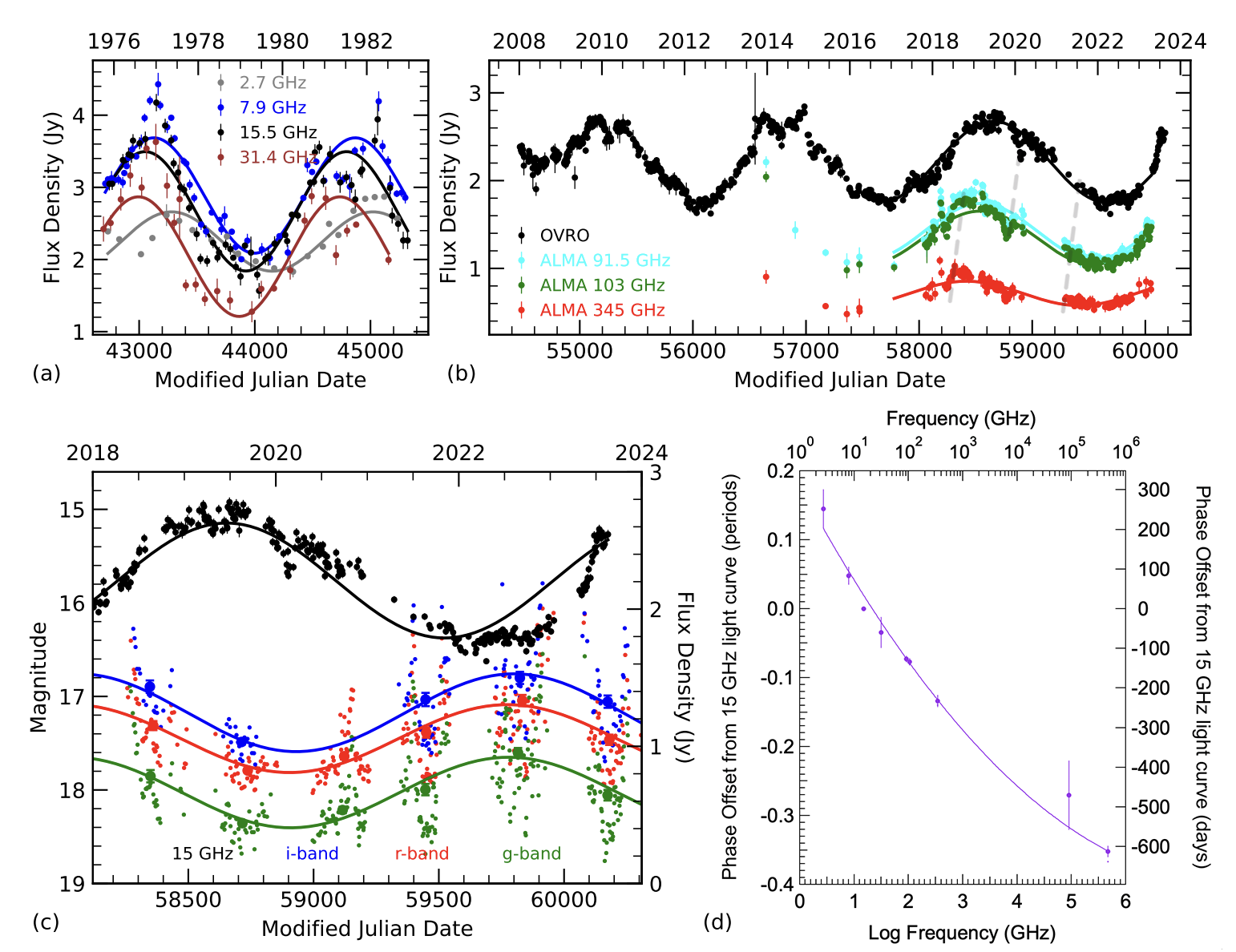}
      \caption{PKS~2131$-$021 light curves at radio, mm, sub-mm,  and optical wavelengths showing a steady monotonic progression of the phase of the sinusoidal variations with frequency. (a)~Haystack 2.7--31.4 GHz observations \citep{1986AJ.....92.1262O} together with the least-squares sine-wave fits to the data at each of the frequencies. There is a steady progression in phase with frequency between the four light curves, with the low-frequency sine waves lagging the high-frequency sine waves. (b)~OVRO 15~GHz light curve plus ALMA light curves at~91.5, 104,  and~345~GHz. As in (a), there is a continuing progression in the phases of the sine wave fits with frequency. The light gray dashed lines connect several Type-A correlated noise variations across various frequency bands (see text). (c) Comparison of the ZTF optical and OVRO radio light curves of PKS~2131$-$021.  The ZTF  $i$, $r$, and $g$-band light curves of PKS~2131$-$021 are shown. The large  symbols are average values for each year.  The black points show the OVRO 15~GHz points over this period, which provide the phase reference point.  The curves show the least squares sine wave fits to the corresponding data. We associate the optical peak with the \textit{nearest} radio peak, not the preceding radio peak.  (d) The observed phase shifts, relative to the 15.5 GHz (Haystack) and  15 GHz (OVRO) light curves, of the sine-wave fits to the data, given in Table \ref{tab:phases}.  A positive phase shift indicates that the light curve is shifted to a later time than the 15 GHz light curve. The curve shows a quadratic polynomial fit to the phase offset \textit{vs.} log(frequency) data (see \S \ref{sec:coherence}). }
         \label{plt:lightcurves3}
\end{figure*}

In Fig. \ref{plt:lightcurves3}(a) and (b) we clearly see non-sinusoidal random variations in the signal on all time-scales. Furthermore, we see clear examples of non-sinusoidal variations that  are correlated across  different observing frequencies (see the dashed lines in Fig. \ref{plt:lightcurves3}b).

When measuring the phase differences of the sinusoidal variations between different frequency bands, it matters how these variations correlate between frequencies. We consider two cases: 
\vskip 3pt
\noindent
Type-A variations are correlated across frequency with the same time delay as the sinusoidal signal. Such variations could be caused, for example, by variations in fueling. Since these behave like the signal itself, they are not random as far as the phase fit is concerned.  In a sinusoidal fit, Type-A variations enter as noise in the phase measurements, but this noise would be strongly correlated between bands, and would mostly cancel. The cancellation would be imperfect if the light curves have only partial overlap when measuring \emph{relative} offsets.
\vskip 3pt
\noindent
Type-B variations are not correlated between frequencies, or correlate with a different phase shift to that of the sinusoidal signal, as might happen due to a shock traveling obliquely across the jet. These contribute directly to the error bars of the relative phase measurements.

Due to the paucity of multi-wavelength light curves of SMBHB blazars over periods of several decades, we cannot reliably distinguish between Type-A and Type-B variations, so it is not possible to carry out a full analysis of the Type-A and Type-B variations, and their effect on the derived sine wave period and its uncertainty. A correlated noise fit as in Appendix~\ref{sec:correlated} would need to be based on how the correlated noise at different frequencies correlate with each other.  A fit based on the correlated noise for each light curve in isolation  greatly overestimates the uncertainty of the relative phase offset between frequencies.

 If the correlated non-sinusoidal variations are dominated by Type-A errors, as appears to be the case, then we can either:

\vskip 1pt
\noindent
1. Fit the lightcurve at each frequency as a scaled and offset version of the OVRO light curve, or

\vskip 1pt
\noindent
2. Fit a phase per light curve using a sinusoidal model with the period fixed at the value for the best fitting OVRO light curve during the period of the overlapping multi-frequency light curves,  assuming white noise. It is important to consider only the period of overlap because of the changes in period with time due to long-term random variations, as described in detail in Paper 1, and as can be seen clearly  in Fig. \ref{plt:lightcurves}(b).  

We adopt alternative 2 in this paper.  Accordingly, we analyzed seperately the following two periods  in the light curves shown in Figs. \ref{plt:lightcurves3} (a) and (b):
\vskip 3pt
\noindent
(i) $45500\;< \: \textrm{MJD} \; 52850$: Here we first carried out the sinusoidal fit to the Haystack 7.9 GHz and 15.5 GHz data, since these are the most heavily sampled Haystack light curves, and we derived a period $P=1742$ days.  We then held this period fixed and carried out the sinusoidal fits to the 2.7 GHz, 7.9 GHz, 15.5 GHz and 31.4 GHz light curves, and measured the phase offsets relative to the phase of the 15.5 GHz light curve at each frequency.  

\vskip 3pt
\noindent
(ii) $57500\;< \: \textrm{MJD} \; 59800$: Here we first carried out the sinusoidal fit to the OVRO 15 GHz data, and we derived a period $P=2002.9$ days.  We then held this period fixed and carried out the sinusoidal fits to the ALMA 91.5 GHz, 104 GHz, and 345 GHz and measured the phase offsets relative to the phase of the 15 GHz light curve at each frequency.  The results of these fits are shown in  Table \ref{tab:phases} and in Fig. \ref{plt:lightcurves3}(d).

\begin{deluxetable}{c@{\hskip 8mm}ccc}[!ht]
\tablecaption{Phase Shift with Frequency Relative to the 15.5~GHz and 15~GHz  Light Curves }
\tablehead{Frequency&Phase&Uncertainty\\
Band&(Fraction&(Fraction\\
&of a cycle)&of a cycle)
}
\startdata
2.7 GHz&0.145&0.028\\
7.9 GHz&0.048&0.013\\
15 and 15.5 GHz&0&0\\
31.4 GHz&$-0.034$&0.022\\
91.5 GHz&$-0.073$&0.005\\
104 GHz&$-0.077$&0.005\\
345 GHz&$-0.134$&0.009\\
infrared&$-0.27$&0.05\\
Optical&$-0.352$&0.008\\
\enddata
\tablecomments{The 15.5~GHz light curve was observed in~1975--1983, and the 15~GHz light curve was observed in~2008--2023), so these were used as the phase reference in both the Haystack data and  the OVRO+ALMA data. The infrared and optical light curves are referred to the 15 GHz OVRO light curve.}
\label{tab:phases}
\end{deluxetable}

\begin{deluxetable*}{c@{\hskip 8mm}ccccccc}[!hbt]
\tablecaption{Probabilities and Significance Levels of GLS Tests from Simulations with Matched Red Noise Tail}
\tablehead{Test&Test& GLS $\mathcal{P}_{\rm peak}$ & Period Range ($\Delta P$)& Total& Number that& $p$-value & Significance\\
Number&&(max=1)&(days)&Simulations&pass test&&($\sigma$)}
\startdata
1& OVRO  &  0.83 &All&$10^6$  &  {53} & {$5.3 \times 10^{-5}$} & {3.88}\\
2&Haystack  &$\ge 0.50$ &1657.5 -- 1854.1&10000  &  197 & $1.97 \times 10^{-2}$ & 2.06\\
3& phase& -&-&-&-&0.2&$0.84$\\
\hline
-& 1+2+3& -&-&-&-&{$2.09\times 10^{-7}$ }& {$5.06$}\\
\enddata
\tablecomments{$\mathcal{P}$ is the GLS power and $\Delta P$ is the range of periods included in the test. The tests are described in Sec.~\ref{sec:statistics}. }
\label{tab:lsvalues}
\end{deluxetable*}

\section{Statistical Significance of the Sinusoid Parameter Agreement in 1975--1981 {\textit vs.} 2008--2023}\label{sec:statistics}

We carried out the generalized Lomb--Scargle (GLS) \citep{Lomb1976,scargle1982,2009A&A...496..577Z} periodogram  analysis described in Section~3.4 of Paper~1. With this analysis we address the question of whether a random process can produce the detected periodic signals in the OVRO and Haystack data. The null hypothesis is that the light curves originate from a power-law red noise process that reproduces the power spectral density (PSD) and probability density function (PDF). We model the PSD with a frequency ($\nu$) dependent power-law $\propto \nu^{-\beta}$ with index $\beta=2.05$. We model the PDF through the empirical cumulative distribution function of the original light curve. We refer to Appendix~A in Paper~1 for details. We devised three tests, the results of which are given in Table \ref{tab:lsvalues}:

Test 1 -- OVRO data: Under the null hypothesis, what is the probability that the strongest peak in the GLS periodogram has a higher test statistic than the GLS peak of the the OVRO data? The test statistic is the significance of the GLS peak estimated at the peak frequency. We refer to Appendix~A of Paper~1 for a detailed description of the procedure. Based on $10^6$ simulations we estimate a probability of {$5.3 \times 10^{-5}$ ($3.88\sigma$)}.

Test 2 -- Haystack data period: Under the null hypothesis and knowing the period in the OVRO data, 
what is the probability that the strongest peak in the GLS periodogram of the Haystack data (a) has a power of at least 0.5 and (b) appears at a period in the $3\sigma$-uncertainty range spanning  the period detected in the OVRO data?
Based on 10,000 simulations we estimate a probability of $1.97 \times 10^{-2}$
($2.06\sigma$).

Test 3 -- Haystack data phase: Under the null hypothesis, and assuming the Haystack data shows a periodicity with period matching that of the OVRO data, what is the probability that the phases match to within the observed 20\% of a half-period offset seen between the Haystack and OVRO sine wave fits? In the case of a randomly produced  periodicity, the phase would be uniformly distributed. Therefore, the probability for test~3 is 0.2 (Paper 1).

We estimate the probability of all three observations as defined in tests 1--3 under the null hypothesis by multiplying the probabilities of the three independent tests. The joint probability is $2.09 \times 10^{-7}$ ($5.06\sigma$).

We reject the null hypothesis at $>3\sigma$ significance level based on the OVRO data alone, and with much higher confidence based on the joint results from the OVRO and Haystack data analysis. It is highly unlikely that a random process produced the observed periodicity. Consequentially, a non-random process must have produced the periodic signal.

These results provide compelling evidence that there is  a long-term persistent periodic physical phenomenon in PKS~2131$-$021 that, given the 19-year gap in the sinusoidal variations from 1984 to 2003, produces intermittent sinusoidal variations in its light curve. In our view it is highly likely, based on the 15 GHz observations alone,  that the variations are due to the orbital motion of a jet-producing SMBH in an SMBHB.

\section{The Radio Spectrum, Structure,  and Polarization of  PKS~2131$-$021}\label{sec:specpol}
\begin{figure}[!b]
   \centering
   \includegraphics[width=0.9\linewidth]{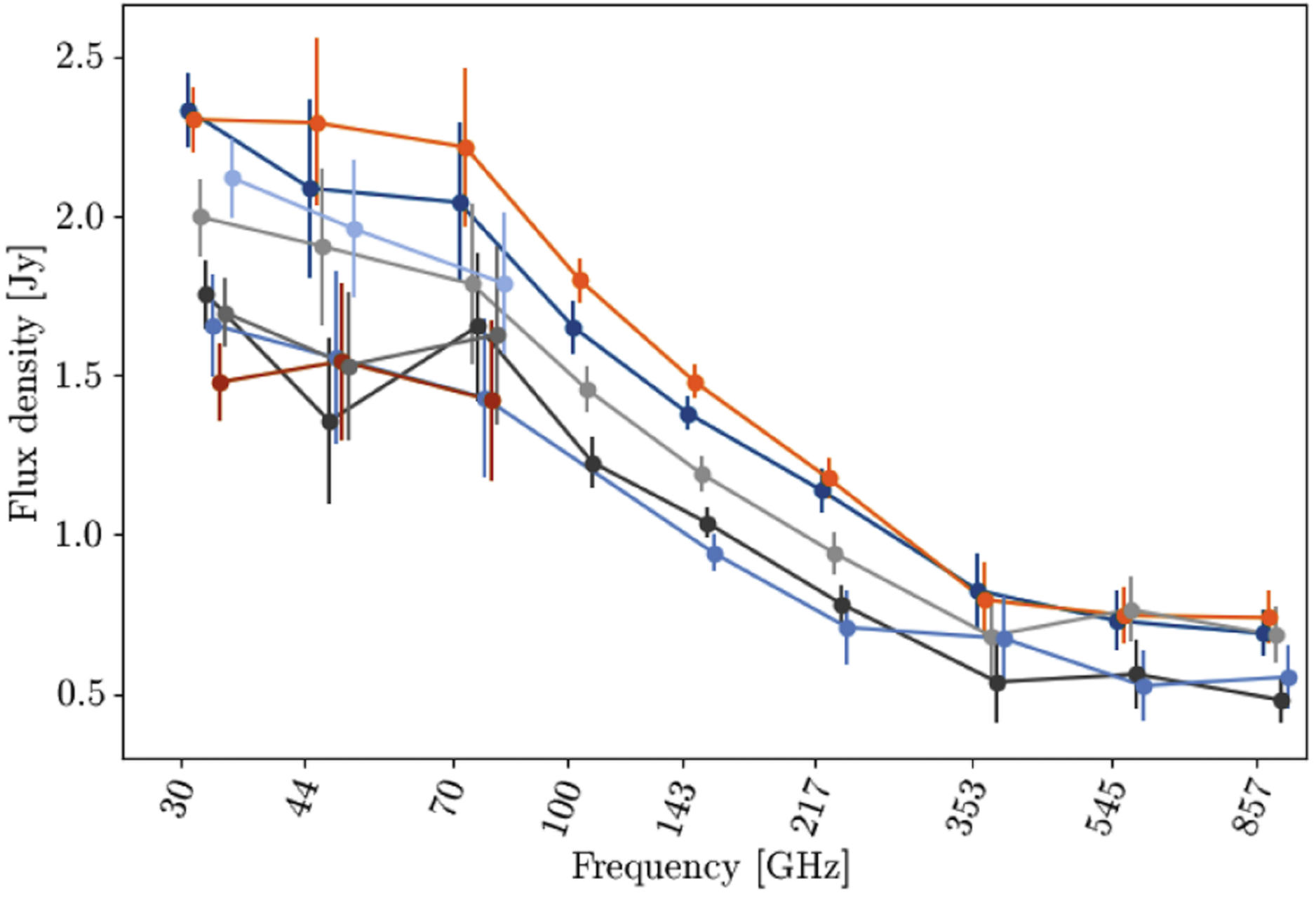}
      \caption{The radio spectrum of PKS~2131$-$021 from centimeter to sub-millimeter wavelengths measured by the Planck observatory at eight different epochs \citep{2023A&A...669A..92R}.}
         \label{plt:planckspec}
\end{figure}

The radio spectrum of PKS~2131$-$021, as measured from centimeter to sub-millimeter wavelengths by the Planck observatory at eight different epochs \citep{2023A&A...669A..92R}, is shown in Fig.~\ref{plt:planckspec}.  Below 30\,GHz the spectrum continues flat all the way down to 100\,MHz (NASA/IPAC NED). It is somewhat unusual for a blazar to have a flat radio spectrum all the way from 100\,MHz to 70\,GHz. This is indicative of a very compact jet structure. Above 70\,GHz, the emission transitions from optically thick to optically thin, and the spectrum steepens to a spectral index $\alpha = -0.74$, and then above 350\,GHz it flattens again---possibly due to the ejection of a new jet component. 

The flat radio spectrum of PKS~2131$-$021 from \hbox{$<100$\,MHz} to 70\,GHz must be due to synchrotron self-absorption \citep{1968ARA&A...6..321S}, with the lower-frequency radiation coming from further out along the jet than the higher-frequency radiation, as is common in blazar  sources  with ``core-jet'' morphologies, such as PKS~2131$-$021 \citep{1978Natur.276..768R,1980IAUS...92..165R}.

The centimeter-wavelength single-dish linear polarization of PKS~2131$-$021 was monitored by the University of Michigan Radio Astronomy Observatory (UMRAO), where the source showed a fairly low typical fractional polarization of 2.8\% at 14.5\,GHz \citep{aller99}. The source was also included in the VLBA polarization monitoring program by \cite{gabuzda00}.

\begin{figure}[!t]
   \centering
   \includegraphics[width=0.9\linewidth]{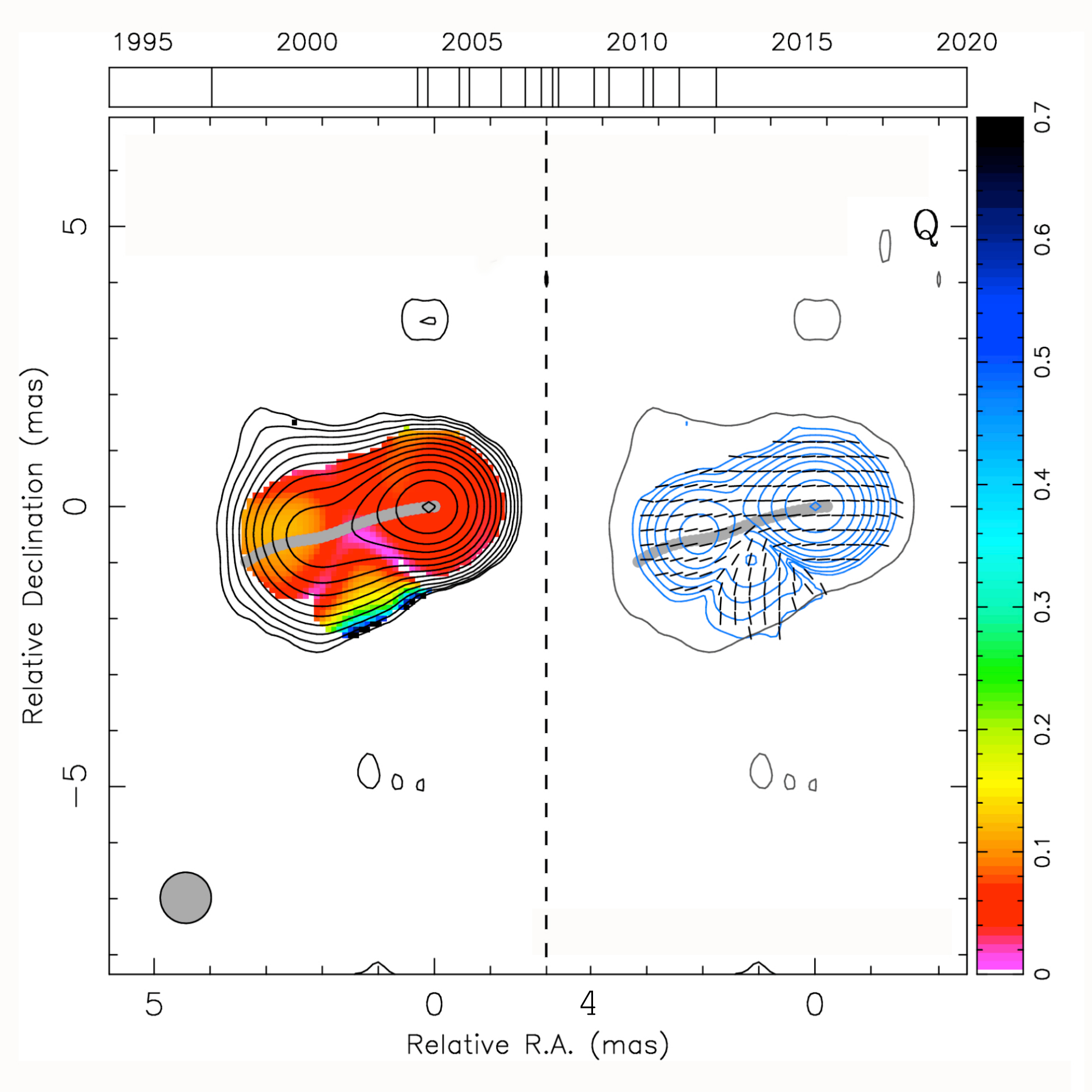}
      \caption{Stacked polarization image of PKS~2131$-$021 at 15 GHz from the MOJAVE program \citep{pushkarev2023}. The left panel shows the stacked total intensity emission over the 16 epochs between 1997 and 2012 in contours, while the color scale indicates the amount of fractional polarization. The right panel shows the lowest total intensity contour (black) and polarized intensity contours (blue) along with the direction of the EVPA (black tick marks). The gray curve shows the ridge line in total intensity.  The timeline on the top of the image indicates the observing epochs by vertical lines.}
         \label{plt:polimage}
\end{figure}

Figure~\ref{plt:polimage} shows the stacked-epoch VLBA total intensity (left panel) and polarization (right panel) images of PKS~2131$-$021 at 15\,GHz from the MOJAVE program \citep{pushkarev2023}. The epochs that have been stacked are shown at the top of the figure. We see that PKS~2131$-$021 has the typical ``core-jet'' structure common to blazars.

The polarization image  of Fig.~\ref{plt:polimage} is produced by first averaging the total intensity $I$ and Stokes parameters $Q$ and $U$ over the 16~observing epochs between 1997 and 2012, and then calculating the fractional polarization and electric vector position angle (EVPA) images from the stacked quantities \citep[see][for details]{pushkarev2023}. Stacking the polarization this way reveals the underlying magnetic field structure. It also reveals a much broader cross-section of the entire jet, which is not often fully visible in individual epochs. 

The EVPA vectors shown on the right in Fig.~\ref{plt:polimage} indicate that the electric field is predominantly parallel to the jet direction, assuming that the Faraday rotation in the jet is small, as was shown in \cite{2012AJ....144..105H}. The only exception to this is the region south-east from the core, which is due to a single component (component~3, see Paper~1), showing a transverse polarization structure. Predominantly parallel EVPA in an optically thin jet, as in this source \citep{2014AJ....147..143H}, means that the underlying magnetic field is dominated by a toroidal component.

\begin{figure}[!t]
   \centering
   \includegraphics[width=1.0\linewidth]{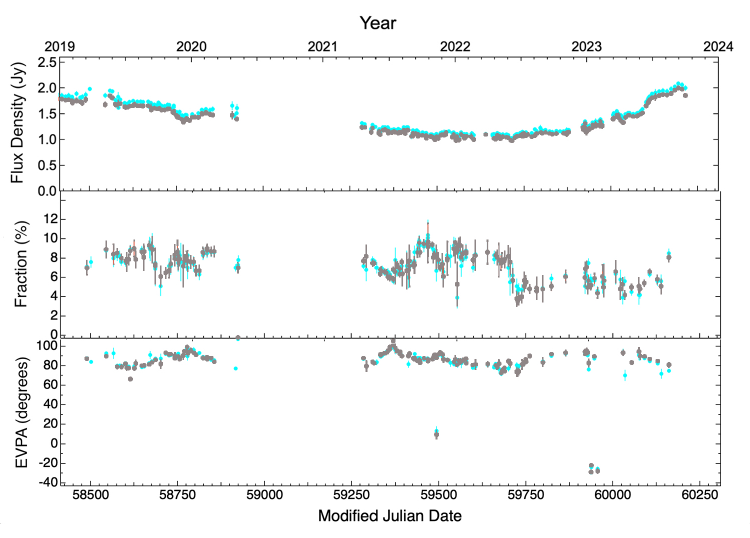}
      \caption{Polarization of PKS~2131$-$021 observed in ALMA Band~3 (Cyan points: 91.5~GHz; gray points: 103.5~GHz.). The EVPA changes little, if at all, between~15~GHz and~345~GHz, and shows no dramatic shift between the optically thick and optically thin regimes.  }
         \label{plt:polarization}
\end{figure}

The integrated linear polarization in PKS~2131$-$021 stays predominantly parallel to the jet direction all the way up to 350\,GHz \citep{2023A&A...669A..92R}. The variation in the linear polarization at 91.5~GHz and 103.5~GHz from 2019 to 2023.5 seen by ALMA is shown in Fig.~\ref{plt:polarization}, where it is seen that the EVPA changes little throughout the sinusoidal cycle. 

These linear polarization observations show that the magnetic field in PKS~2131$-$021 is dominated by the helical component $B_\phi$, and that there must be a strong  current, $I$, along the jet.

Circular polarization observations of PKS~2131$-$021 are also available from the MOJAVE program. In the first epoch of MOJAVE observations in May 2003, no significant circular polarization was detected; the upper limit on the fractional circular polarization was $m_c < 0.19$\% \citep{homan2006}. This is typical behavior at 15\,GHz, where only 17 out of 133 sources in the single-epoch study of \citet{homan2006} showed significant circular polarization detections at higher than $3\sigma$ level. However, in a multi-epoch study covering all MOJAVE data until the end of 2009, the source is detected  in circular polarization with $m_c=(-0.456\pm0.095)$\% during one epoch in February 2009 \citep{homan2018}. During this epoch, the linear polarization of the core component reaches its highest level $m_l=7.0$\% within the MOJAVE epochs analyzed until the end of 2009. As discussed in \citet{homan2018}, it is not unusual for a source to show significant circular polarization at single epochs, and 91 out of the 278 sources showed a significant detection at least during one observing epoch, with typical detection levels being in the range 0.3--0.7\%. The source was also observed at 15, 22, and 43\,GHz by \cite{vitrishchak08}, but only upper limits on circular polarization were derived with limits of $< 0.24$\%, $<0.70$\%, and $0.78$\% at 15, 22, and 43\,GHz, respectively.

Circular polarization can be  due either to intrinsic circular polarization of the synchrotron emission or to Faraday conversion of linear to circular polarization \citep[e.g.,][]{jones1977}. At centimeter wavelengths, it is often attributed to Faraday conversion \citep[e.g.,][]{osullivan2013}, although in the inhomogeneous VLBI cores it can also be due to intrinsic circular polarization \citep{homan2009}. With only  single frequency observations it is impossible to distinguish between the two scenarios.

 \section{The Centimeter-to-Optical Emission in PKS~2131$-$021}\label{sec:multifrequency}

  In Fig. \ref{plt:lightcurves3} (a), (b), and (d), we saw that the phase shift varies monotonically with frequency from 2.7 GHz to 345 GHz. In this section we show that this coherent  phase shift with frequency continues through the infrared to the optical band.  To the best of our knowledge, such coherence in the flux density variations in a blazar, extending from the radio band to the optical band, has not been observed before. We suggest that this phase shift with frequency is due to optical depth effects in the jet.

 \subsection{The 2.8$\,\mu$\lowercase{m}--5.2$\,\mu$\lowercase{m} emission}\label{sec:wise}
 
 We have extracted the Wide Field Infrared Explorer (WISE) data in the 2.8--3.8 micron and 4.1--5.2 micron bands.
In Paper~1 we carried out a cross-correlation analysis of the WISE 1 (2.8--3.8$\mu$m) and WISE 2 (4.1--5.2$\mu$m) data with the OVRO 15~GHz light curve and found marginal $(\sim 2\sigma)$ evidence for a correlation.  The WISE data are very sparsely sampled in time and in our  cross-correlation analysis of the complete data set the peak cross correlation significance is less than $2\sigma$. However we find that the phase of the WISE 1--OVRO cross-correlation data, relative to the 15 GHz light curves, is $-470_{-147}^{+166}$ days, and that of the WISE 2 data is $-470_{-162}^{+177}$ days (i.e., $-0.27_{-0.09}^{+0.10}$ of the period in both cases). 

 The cross-correlation between the OVRO and WISE data was performed following \cite{2014MNRAS.445..437M}. The uncertainty in the delay is derived using 1000 iterations of `random subset selection' and `flux randomization' as described in \cite{1998PASP..110..660P}. Since the WISE data are so sparse, it was necessary to fit the position of the cross-correlation peak because it was not always well defined and showed large fluctuations. The fit was done using a parabola and included the cross-correlation delays from $-1000$ to 0 days. The quoted intervals correspond to 1$\sigma$ error bars. 

The data we have suggest, therefore,  that the sinusoidal signal is present  in the infrared WISE bands.

\subsection{The Strong Optical Sinusoidal Emission}\label{sec:optical}

In Paper~1 we reported that the optical variations in PKS~2131$-$021 observed with the Zwicky Transient Facility  (ZTF; \citealt{2019PASP..131g8001G,2019PASP..131a8003M}) were not correlated  with the 15~GHz variations. The optical data in Paper~1 extended to 2021 Oct 1 and spanned only about half of a single period in the sinusoidal variations. We have now added another two years of ZTF data up to 2023 Sep 9, so that, all told, the ZTF data now span a full period.  We now find that there is a \textit{strong} sinusoidal variation at optical wavelengths of the same frequency as that in the 15 GHz  light curve. 

The ZTF  $g$, $r$, and $i$ band light curves of PKS~2131$-$021 (Fig.~\ref{plt:lightcurves3}c) show a periodic variation having  the same period, to within the uncertainties, as the 15 GHz light curve.  Sine-wave fits to the ZTF data are presented in Appendix~\ref{sec:sineoptical}. where we show that the combined sine-wave fit to all three data sets yields a period of $P_\textrm{optical}=(1764  \pm 36)$ days. This agrees, within the uncertainties, with the period of $(1739.2\pm1.2)$ days determined  from the OVRO+Haystack data. It appears, therefore, that the sinusoidal variations persist at least up to optical wavelengths.  We present evidence below that the same periodicity  likely continues up to $\gamma$-ray energies. 

\begin{figure*}
   \centering
   \includegraphics[width=1.0\linewidth]{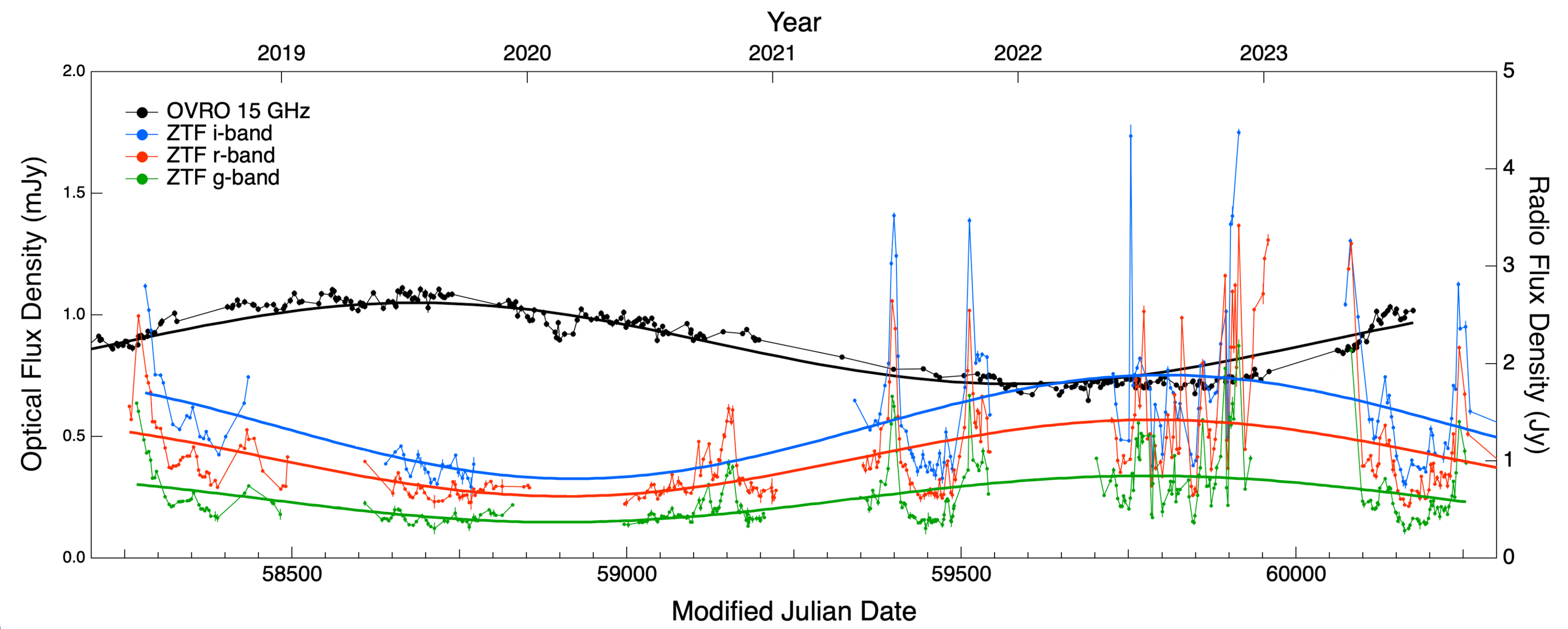}
      \caption{Comparison of the short-term random variations in PKS~2131$-$021 at radio and optical wavelengths \textit{vs.} the sinusoidal variations in these two windows.  The flux density scales have been adjusted such that the amplitudes of the radio and optical sine waves are approximately equal, in order to be able to compare the amplitudes of the short term variations at optical and radio frequencies.  In general the radio short-term variations are significantly smaller than those at optical wavelengths (see text). }
         \label{plt:fluxcomparisons}
\end{figure*}

To determine the phase shift of the optical light curves relative to the OVRO 15 GHz light curve we have used the period $P=1739.2$ days derived for the Haystack 15.5 GHz and OVRO 15 GHz light curves in \S \ref{sec:period-fit}. The reasons for this choice are (i) the period of the optical sine wave is well determined, as can be seen from Fig. \ref{plt:lightcurves3}(c),  (ii) it is consistent with the period derived from the OVRO+Haystack light curves, and furthermore (iii) in Fig. \ref{plt:fluxcomparisons}, where we compare the OVRO 15 GHz light curve with the ZTF light curves, it is immediately clear that the non-sinusoidal variations in the optical light curves are not seen at 15 GHz.   The derived value of the optical sinusoidal shift is given in Table \ref{tab:phases}.

The time shift between the observed maxima in the sine wave fits to the optical light curves and the  15 GHz curve is  612 or 1127 days, depending on the direction of the shift. The more obvious choice is 612 days, since this shifts the light curve in the same direction relative to the 15 GHz OVRO light curve as we determined from cm to sub-mm wavelengths, and continues the monotonic trend we have observed at the lower frequencies (see Fig.~\ref{plt:lightcurves3}(d)).  We return to this point in section \ref{sec:gamma}.

\subsection{The Comparative Difficulty of Detecting Optical \textit{vs.} Radio Sinusoidal Variations in Blazars}\label{sec:comparative}

This is the first detection of statistically compelling sinusoidal variations in a blazar at both optical and radio wavelengths. It provides the opportunity to estimate the relative difficulty of making such detections in these two windows.  As shown in Paper~1, the peak random fractional variations in the radio light curve are $\sim 20\%$ (see Fig.~10(c) of Paper~1).  From  the ZTF data of  Figs.~\ref{plt:lightcurves3}(c)  we find that the peak random fractional variations in the optical light curve are $\sim 200\%$. This order of magnitude difference, coupled with the extraordinary challenges of determining the high statistical significance of the sinusoidal variations in PKS~2131$-$021, with all the various types of confusion that are introduced by the random variations  discussed in Paper~1 and the inevitable annual gaps in the optical light curves, reveals that the detection of statistically significant optical sinusoidal variations in blazars is a daunting challenge. The situation is illustrated in Fig.~\ref{plt:fluxcomparisons}. There are periods when the optical short-term variations are no larger than
those at radio wavelengths, but there are also periods when
they are an order of magnitude,
or more, greater than those at radio wavelengths, making it
more difficult to detect sinusoidal variations at
optical than at radio wavelengths.  

It is clear that, in optical searches for sinusoidal blazar variations, a minimum of \textit{both} $\sim100$ observations per year \textit{and} rigorous statistical red noise tests for global significance are needed to detect sinusoidal variability in blazars with periods of several years. Without such dense sampling and rigorous statistical testing there is a high probability of spurious detections of sinusoidal periodicities in blazars at \textit{both} radio \textit{and} optical wavelengths.

In order to obtain a qualitative measure of the relative difficulty of detecting a blazar showing sinusoidal  variations like those seen in PKS~2131$-$021 at optical and radio wavelengths, we have carried out the same GLS analysis on the optical data as on the radio data, truncated to the ZTF time frame, using the optical data shown in Fig.~\ref{plt:fluxcomparisons}. Based on 20,000 simulations, we found that the global $p$-values of the GLS peaks in the $i$, $r$, and $g$ bands are 0.5, 0.4, and 0.5, respectively, whereas that at 15 GHz is 0.002.

\subsection{Coherence of the Periodicity from the Radio to the Optical Emission}\label{sec:coherence}

The phase shifts measured in the previous sections are tabulated in Table \ref{tab:phases} and plotted in Fig.~\ref{plt:lightcurves3}(d).  They are well-fitted by a linear-logarithmic fit, as shown on the linear-log plot of Fig.~\ref{plt:lightcurves3}(d). A linear fit to the  2.7 GHz - 345 GHz data is given by 
\begin{equation}
\delta \phi = 0.020 -0.116 \log_{10}(\nu/15)
\end{equation}
where $\nu$ is in GHz.  We make use of this relationship in equation (\ref{eqn:rnu}) below.  The curved line of Fig.\ \ref{plt:lightcurves3}(d) shows a quadratic polynomial fit to all the data given by $\delta \phi = 0.178 -0.146\; \log_{10}(\nu) + 0.0093[\log_{10}(\nu)]^2$.

The monotonic variation in phase lag with frequency from 2.7 GHz to optical frequencies demonstrates convincingly that the sinusoidal variation is caused by a physical process in the source. 

\begin{figure}[!b]
   \centering
   \includegraphics[width=1.0\linewidth]{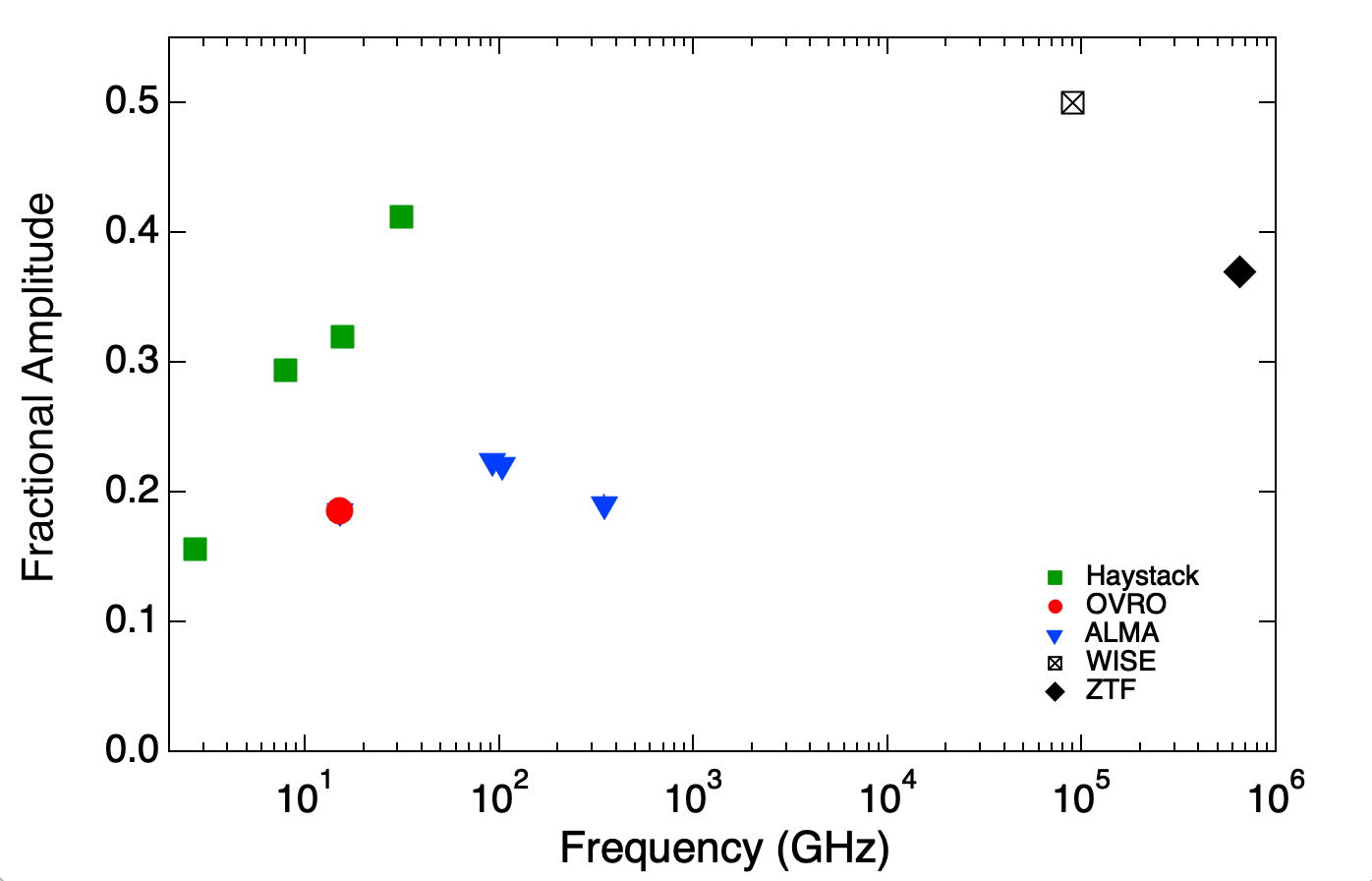}
      \caption{ Fractional amplitudes of the fitted sine waves as a function of frequency. The amplitudes are normalized by the mean amplitude of each light curve. The data displayed are:  Haystack data, from 1975 to 1983 (green squares), OVRO (red circle) and ALMA  (blue points) data from 2014 to 2023, the ZTF data (black diamond),  and the measured fractional amplitude of the variations in the  WISE data (black crossed square).  In all cases the error bars are smaller than the symbols. }
         \label{plt:fraction}
\end{figure}

\subsection{The Fractional Amplitude of the Sinusoidal Variations}\label{sec:fractional}
 In Fig.~\ref{plt:fraction} we show the fractional amplitudes  of the sine wave fits to the Haystack, OVRO, ALMA, WISE and ZTF  data (i.e., the sine-wave amplitudes divided by the mean flux).  The fractional amplitude of the sine wave observed between 1975 and 1983 at Haystack is clearly dropping at the lowest frequency (2.7 GHz).   The blue points (OVRO+ALMA) and the black point (ZTF) have considerably more data than the green points (Haystack) and the WISE data.

\begin{figure}[!b]
   \centering
   \includegraphics[width=1.0\linewidth]{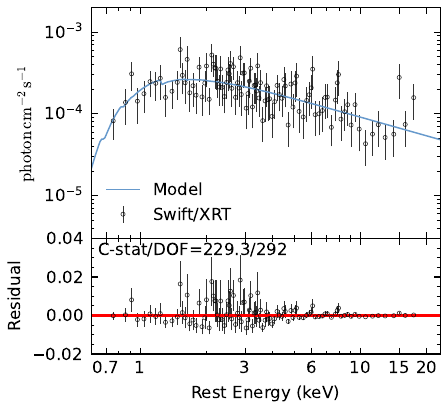}
      \caption{Stacked \textit{Swift/XRT} spectrum of PKS~2131$-$021 shifted to the rest frame of $z=1.283$. The spectrum is characterized by an absorbed powerlaw with photon index $\Gamma=1.78\pm0.15$ and rest frame hydrogen column density $N_\mathrm{H}=3.4^{+2.7}_{-2.4}\times10^{21}\,\rm cm^{-2}$ (at $z=1.283$). Data have been visually rebinned to S/N$>$2.
      } 
         \label{plt:swift-spectrum}
\end{figure}

\section{X-ray Observations and Future Prospects  }\label{sec:x-ray}

Hard X-ray emission from blazars is important because it is directly connected to the central engine \citep[scale being smaller than gravitational radius,][]{Bhatta2018A&A}. The soft X-ray emission from blazars is thought to be due to synchrotron emission, while the hard X-ray emission is thought to be caused by up-scattering of soft photons either from the synchrotron emission  \citep{Mastichiadis2002PASA}, or from relatively cool structures like the accretion disk and the broad line region \citep{Dermer1993ApJ...416..458D,Sikora1994ApJS...90..923S}. While significant periodicities are common in the light curves of stellar mass black hole X-ray binary systems, they are surprisingly rare in AGN X-ray observations (however, see \citealt{Jurysek2022icrc}). Only very recently have significant periodic signals started to be widely reported and  studied systematically \citep{Arcodia2021Natur,MiniuttiNatur2019}. In most cases, they originate in stellar-mass compact objects interacting with the AGN accretion disk \citep{Franchini2023}. In the particular case of blazars, the X-ray light curves are usually chaotic and aperiodic. X-ray spectral-timing studies have revealed the existence of time-dependent flux density and spectral state patterns. \citet{Bhatta2018A&A} explored the connections between blazar X-ray variability and other properties, such as spectral hardness and intrinsic X-ray flux in the  \textit{NuSTAR} band. All 13 sources in their sample displayed high-amplitude, rapid, aperiodic variability with a timescale of a few hours.

Since X-rays probe the region close to the central engine, the presence of a close-separation SMBHB could potentially imprint the high-energy emission. Simulations predict that close-separation accreting SMBHBs will have different X-ray spectra to those of single accreting SMBHs \citep{Tang2018MNRAS,Krolik2019ApJ}. Accreting SMBHBs may show a periodically modulated hard X-ray component whose period is of order the binary orbital period, although these predictions have been challenged \citep[e.g., see][]{Saade2023AAS,Saade2024}. There are no available high-quality X-ray data for PKS~2131$-$021, but short exposures during an \textit{XMM-Newton} slew indicate a fairly high X-ray flux ($\sim2\times10^{-12}\rm~erg\,s^{-1}\,cm^{-2}$) in the 2--12\,keV range. We analyzed all available \textit{Swift/XRT} obervations. The stacked spectrum shown in Fig.~\ref{plt:swift-spectrum} reaches a total net exposure time of 20.7\,ks, characterized by a featureless power-law ($\Gamma=1.78\pm0.15$) with a moderate intrinsic hydrogen column density of $N_\mathrm{H}=3.4^{+2.7}_{-2.4}\times10^{21}\,\rm cm^{-2}$ at the rest-frame of $z = 1.283$. We estimate the 0.3--10\,keV unabsorbed flux in the observed frame as $(1.86\pm0.15)\times10^{-12}\,\rm erg\,s^{-1}\,cm^{-2}$. In the near future, long-term monitoring with large-field-of-view instruments such as \textit{eRosita} and \textit{Swift} will be practical and promising, given the relatively high redshift of PKS~2131$-$021.

\section{The $\gamma$-ray Emission}\label{sec:gamma}

We used the Fermi light curve repository\footnote{\url{https://fermi.gsfc.nasa.gov/ssc/data/access/lat/LightCurveRepository/}} \citep{Abdollahi2023} to extract the 30-day binned $\gamma$-ray light curve in the 0.1--100~GeV band. The light curve was generated with the spectral index as a free parameter. We used the test statistic (TS) to identify, by eye, periods with increased significance indicating the presence of increased flux. We then identified the highest flux point in $\gamma$-rays which we match with the brightest radio flares by shifting the $\gamma$-ray data in time.  We assumed the commonly accepted scenario in which the $\gamma$-rays come from a region upstream from the radio emission region \cite[e.g.,][]{Liodakis2018}. 

\begin{figure}
   \centering
   \includegraphics[width=0.9\linewidth]{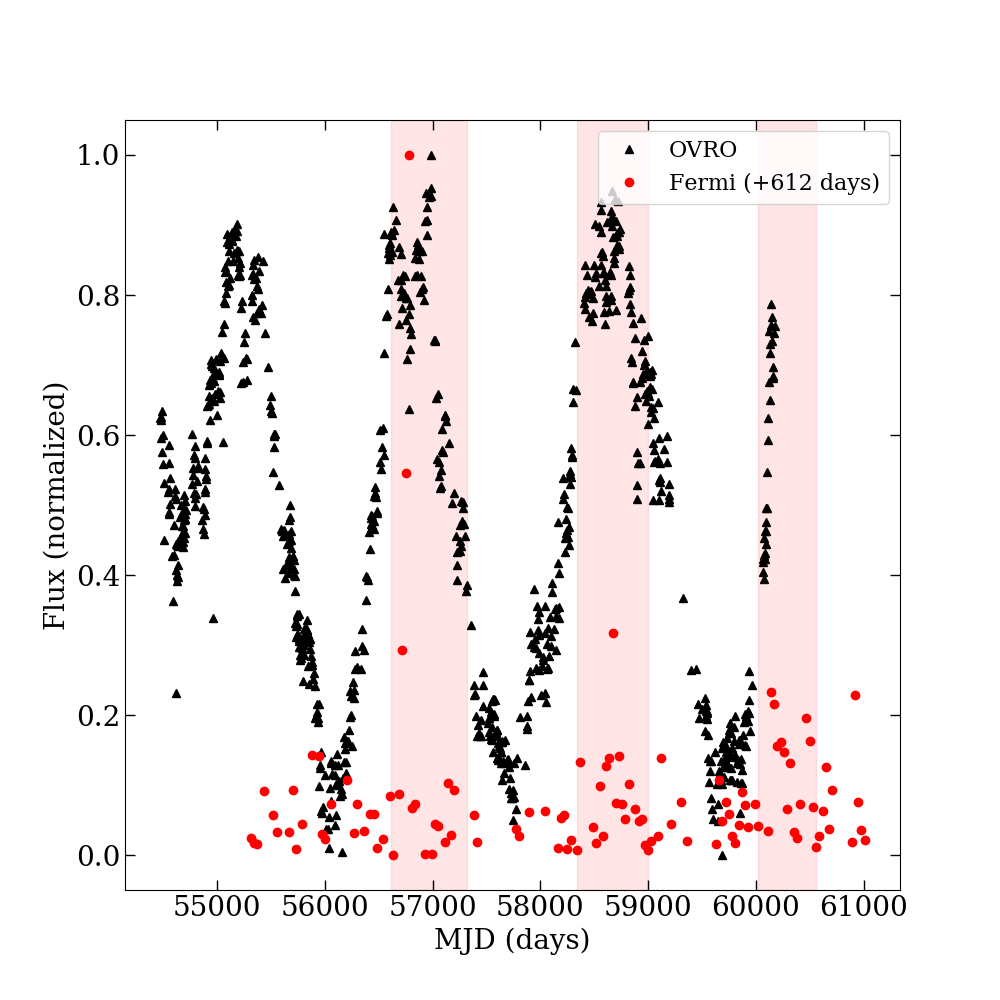}
      \caption{Normalized OVRO and Fermi light curves.  The red shaded areas indicate the extent of the increased $\gamma$-ray activity. The $\gamma$-ray data have been shifted in time by 612 days to align the peaks in the radio and $\gamma$-ray observations (see text).}
         \label{plt:fermi}
\end{figure}

Although $\gamma$-rays show sharper flares compared to the broad radio peaks, it is clear that the enhanced $\gamma$-ray activity periods, when shifted,  align very well with the radio flares (Fig.~\ref{plt:fermi}). Depending on the radio peak, we can achieve a good alignment between radio and $\gamma$-rays with time-lags ranging between 454 and 819 days. If we shift the $\gamma$-rays forward by 612 days to match the radio-optical phase lag we obtain the comparison shown in  Fig.~\ref{plt:fermi}, which appears to us a reasonable match, given the uncertainties. This is consistent with the monotonic phase trend shown in Fig.~\ref{plt:lightcurves3}(d), which includes the time-lag found between radio and the WISE observations of 470 days (see Section \ref{sec:wise}). 
The fact that the radio--$\gamma$-ray time-lags are consistent with what we find for the infrared and optical observations suggests inverse-Compton scattering for the $\gamma$-ray production mechanism \cite[e.g.,][]{Liodakis2019}.

\section{Kinematic Orbital Model}\label{sec:model}

In this section we consider the implications of the new observations for the KO model, and explore the physical properties of PKS~2131$-$021 within the context of this model. 

\begin{figure}[!t]
   \centering
   \includegraphics[width=0.9\linewidth]{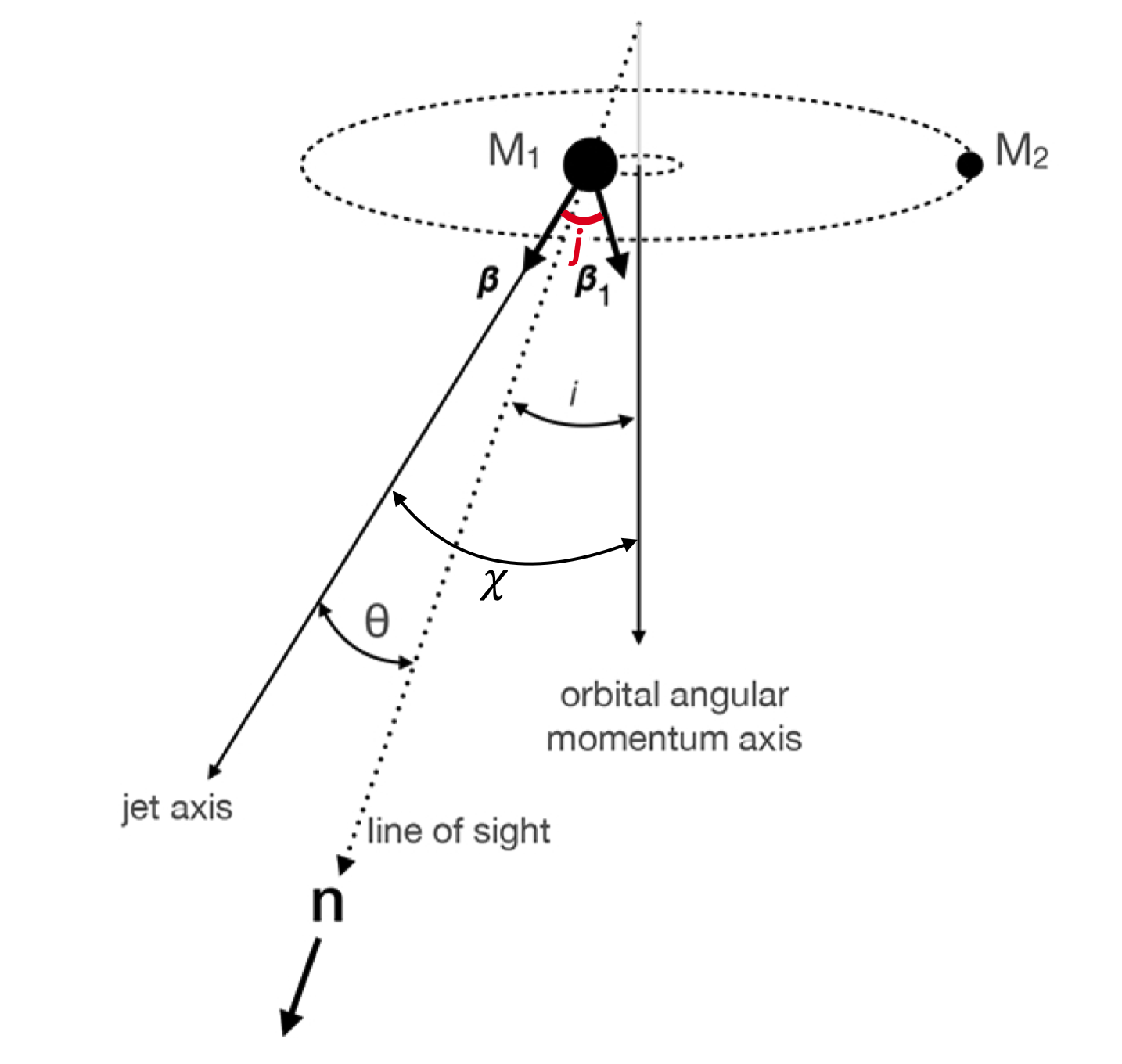}
      \caption{ The KO model. In Paper~1 it was assumed that the angle between the jet axis, $\bm{\beta}$, and the orbital angular velocity, $\bm{\beta}_1$, is $j\sim \pi/2$, leading to equation (8)  in Paper~1. In this figure $j$ is indicated in red.}
         \label{plt:model}
\end{figure}

\begin{figure*}[!t]
   \centering
   \includegraphics[width=1\linewidth]{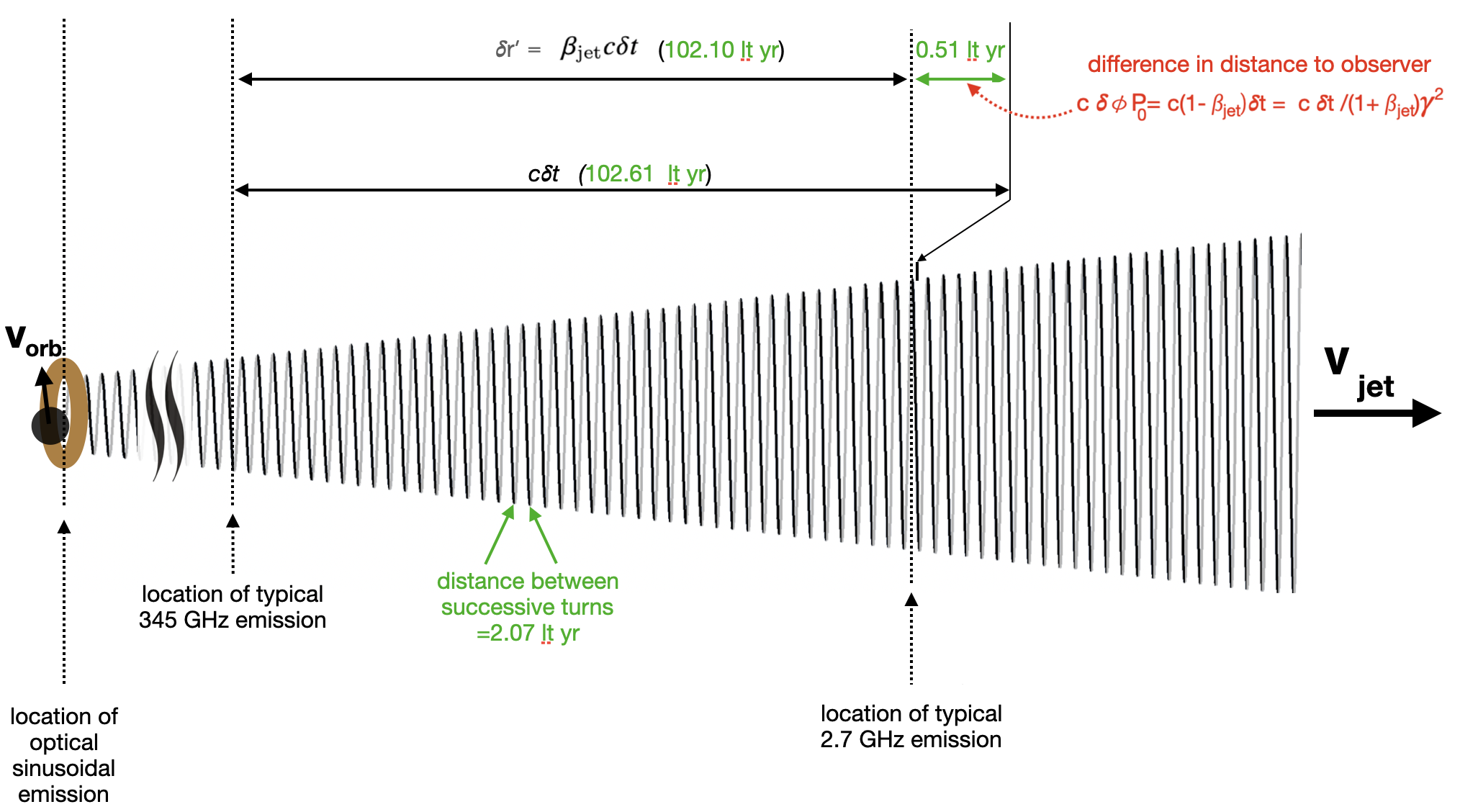}
   \caption{Schematic diagram (not to scale) showing the relationship between the typical locations of the emission at 345 GHz and 2.7 GHz on the KO model, for the case $\beta_1=0.020$,  $\beta  =0.995$, and $\gamma\sim10$. The numbers in green script correspond to this case. On this model we assume that the dominant emission at each frequency arises from the region where the jet becomes optically thick at that frequency. This schematic illustrates the special case where the orbital plane  and the jet axis are orthogonal, ($j=\pi/2$).}
   \label{plt:cartoon}
\end{figure*}

A schematic of the KO model is shown in  Fig.~\ref{plt:model}. The model consists  of a black hole binary. Expressing  the masses of the binary components in units of $10^8M_\odot$ as $M_{i\,8},\, i=1,2$, the primary mass is  $M_1=10^8M_{1\,8}M_{\odot}$, and the secondary mass is $M_2=10^8M_{2\,8}M_{\odot}$. Both masses orbit the center of mass with period  in the rest frame of the binary $P_0\!=\!761.8\,{\rm d}$, and angular momentum that makes an angle $i$ with the line-of-sight unit vector {\bf n}. We assume that the motion has been circularized by frictional drag \citep{1980Natur.287..307B}, but elliptical orbits also work. We assume that the jet is launched along the spin axis of $M_1$ with fixed velocity $c{\bm \beta}$ relative to the black hole.  We know from the  VLBI observations discussed in Appendix~\ref{sec:brightness} that the line of sight is inclined at a  small angle, $\theta$, to the jet axis.  The relationship between the orbital velocity of $M_1$ and ${\beta}_{\rm  1}$  is  ${\beta}_{\rm  1}=0.036M_{2\,8}/(M_{1\,8}+M_{2\,8})^{2/3}$. Here we have set $c=1$, and we do so for the rest of this section.

The orbital motion changes the velocity of the emitting material in the jet relative to the observer, and hence the Doppler factor and beaming. In the case of PKS~2131$-$021 the Doppler factor of the jet is high, so that the orbital motion can have a significant effect on the Doppler factor.

Suppose we have a source emitting isotropically in its rest frame with a flux density $S'$. The observed flux density will be given by $S= {\cal D} ^{2-\alpha}S'$ \citep{1979Natur.277..182S}, where $\alpha=d\ln S/d\ln\nu$ is the spectral index, and ${\cal D}$ is the Doppler factor.

In this paper, as in Paper~1, in order to maximize the amplitude of the sinusoidal wave, we assume that the angle between the normal to the orbital plane and the line of sight, $i$, is $\sim 0$. 

\subsection{Consistency of the observed light curves with the Kinematic Orbital Model }\label{sec:consistency}

In Appendix~\ref{sec:brightness}, we show, based on MOJAVE VLBI measurements, that the  Doppler factor  of the core component in PKS~2131$-$021 is $\cal{D}=14$, which, using the angle between the jet axis and the line of sight of $\theta=3.8^\circ$ \citep{2021ApJ...923...67H}, yields a Lorentz factor ${\rm \gamma} = 10$  (see Appendix \ref{sec:brightness}).   We can perform a simple consistency check with the KO model  using the phase shifts we have determined.  This also provides some useful constraints on the scales of the region producing the sinusoid in PKS~2131$-$021.

\begin{figure*}
   \centering
   \includegraphics[width=1.0\linewidth]{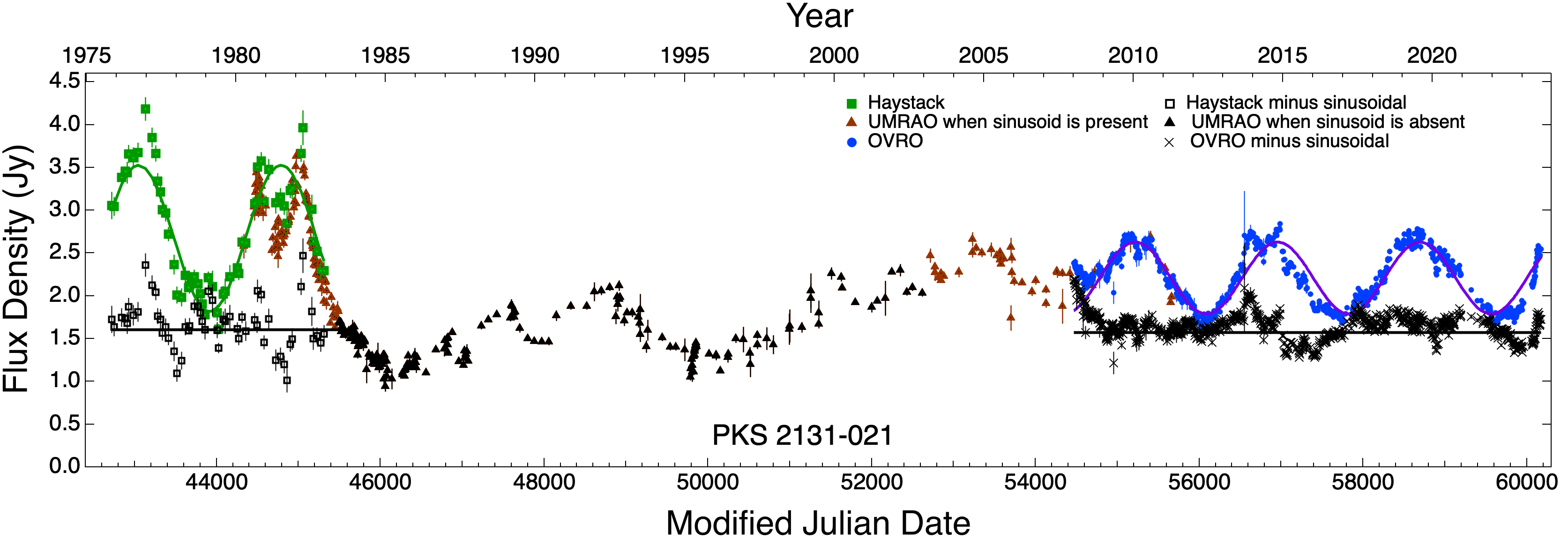}
      \caption{Sinusoidally-varying and non-sinusoidal-varying components in PKS~2131$-$021. Green squares: Haystack 15.5 GHz observations \citep{1986AJ.....92.1262O}, together with the least-squares sine-wave fit (green line) to the data from Fig.~\ref{plt:lightcurves3}(a). Black and brown triangles: UMRAO data \citep{1985ApJS...59..513A,2017Galax...5...75A}. Blue circles: OVRO 15~GHz data, together with the best OVRO least squares sine wave fit (purple line)  from Fig.~\ref{plt:lightcurves}(a), the parameters of which are given in Table \ref{tab:bestfit}. The black horizontal lines show the average level of the non-sinusoidally-varying component corresponding to a fractional variation of 0.72 in the sinusoidally varying components (see text). The black open square (cross)  points show the Haystack (OVRO) residual non-sinusoidally-varying flux density after subtraction of the corresponding sine wave assuming this fractional variation of 0.72.  Thus the black data points across the whole span of time show the light curve of PKS~2131$-$021 as it would appear without the sinusoidally varying component. }
         \label{plt:baselevels}
\end{figure*}
 
We assume that the emission of  observed frequency $\nu$ comes from jet radius $r'(\nu)$ in the SMBHB rest frame.  
Taking into account light travel-time effects (see Fig.~\ref{plt:cartoon}), the difference in $r'(\nu)$  between two phases observed simultaneously (assuming a head-on jet and $\beta \sim 1$) is

\begin{equation}
 \delta r'(\nu) = \beta  (1+\beta  )\gamma^2 c P_0 \delta \phi \sim  2 \gamma^2 c P_0 \delta \phi,
  \label{eqn:deltar}
\end{equation}
where $P_0$ is the binary period in the rest frame of the host galaxy.   Thus the difference in the  distances of the emission sites at two frequencies, $\nu_1$ and  $\nu_2$, is
\begin{equation}
r'(\nu_2) - r'(\nu_1) = -0.232 \gamma^2 c P_0 \log_{10}(\nu_2 / \nu_1),
 \label{eqn:rnu}
\end{equation}
which implies that the frequency of the corresponding peaks in the sinusoidal variations as a function of radius satisfies
\begin{equation}
\nu = \nu_0 \exp[- r' / (0.064 \gamma^2 \, {\rm pc}) ].
  \label{eqn:nu}
\end{equation}

Note that we cannot determine $\nu_0$ with existing data because we only have phase differences.  However, the fact that we determined this linear relation over a factor of 128 in $\nu$ implies that we are seeing at least 5 $e$-folding distances, so we can put a lower limit on $r'$
\begin{equation}
r' > 5 \times 0.064 \gamma^2 {\rm pc} \sim 0.320 \gamma^2 \, {\rm pc} \, ,
  \label{eqn:rdash}
\end{equation}
\noindent
so that, based on the value of $\gamma=10.0125\;(\beta=0.995)$, 
\begin{equation}
r' \gtrsim 32.1 \,{\rm pc, \,i.e.,\, \sim 105 \, ly}.
  \label{eqn:rdash2}
\end{equation}
This also gives a lower limit on the projected scale,
\begin{equation}
r'_{\rm proj} = \theta r' > 0.320 \gamma^2 \theta \, {\rm pc}
  \label{eqn:rdashproj}
\end{equation}
i.e., 
\begin{equation}
r'_{\rm proj} \gtrsim 2.13 \,{\rm pc, i.e.\,}  6.9 \; {\rm ly},
\label{eqn:rdashproj2}
\end{equation}
based on the value of $\theta=3.8^\circ$ derived by \citet{2021ApJ...923...67H}.

In the light curve of PKS~2131$-$021 shown in Fig.~\ref{plt:baselevels}, it can be seen that during the period when the sinusoidal signal was absent, from 1984 to 2003, the flux density varied between 1.0\,Jy and 2.2\,Jy.  The horizontal black lines show the likely mean levels of the non-sinusoidal signal during the epochs when the sinusoidal signal was present.  Any mean levels significantly different from these would lead to unlikely steps in the flux density level at the transition. We use these mean levels to determine the fractional change, $\delta S/S$, in flux density of the sinusoidal signal to be 0.72.  The black points show the sinusoidal-signal-subtracted residual light curves.  The smooth continuation of the light curves across the whole period (black points) is a result of the careful balancing we have done of the fractional change in flux density due to the sine wave and the mean level of the non-sinusoidal component, and shows that this balancing has been done correctly.

This fitting was done by eye, and it was done independently for the transitions at the end of the green sine wave and the beginning of the blue sine wave shown in Fig.~\ref{plt:baselevels}.  The mean levels without the sine waves, shown by the black horizontal lines differ by 2\%, even though these fits were determined at transitions that are separated by 24\,yr. We also find that the mean level during the period when the sinusoidal variation was absent, marked by the black triangles, differs by only 7\% from the mean of the residuals during the two periods when the sinusoidal variations were present. The good agreement between these mean levels lends credence to our hypothesis that there are underlying emission components that are not varying sinusoidally, represented by the black symbols in Fig.~\ref{plt:baselevels}, and that the sinusoidal variations are seen as \textit{an addition} to this underlying component.  

We have not attempted here to subtract the sinusoidal signal during the period from 2003 to 2008, when it appears to have been switching on, because the amplitude was clearly changing. Note that we the fractional variation in flux density of 0.72 is consistent with a value of $\beta_1=0.020$, and $\beta  =0.995$, i.e. $\gamma\sim 10$.    For convenience we will adopt the value $\beta_1=0.020$. 

As an instructive exercise and guide, we consider  the KO model in the rest frame of the SMBHB,  in the case where the orbital plane  and the jet axis are orthogonal. In this case the  locus of the jet is a helix on the surface of  a cone, as shown in Fig.~\ref{plt:cartoon}, and the opening angle of the cone is $2 \beta_1/\beta  $. From Equation~\ref{eqn:deltar} we see that with these parameters $\delta r'$ between the regions emitting at 2.7 GHz and 345 GHz is 116.32\,ly, and the distance along the jet axis between successive turns of the helix is 2.07\,ly.  Thus there are $\sim 60$ turns of the helix between these two surfaces.  At first sight it may seem strange that the phase difference between these two emission regions separated by over 60 cycles is so small, but this is simply a result of the fact that the jet is advancing at nearly the speed of light ($0.995  c$), so that for any distance light travels along the jet axis, the jet falls behind by only 1/200 of that distance, and thus, in $\sim$60 rotations the jet falls behind by only 0.3 of a cycle.  The surface from which the emission emerges at a given frequency is at a fixed distance from the SMBH, but the jet is moving through this stationary surface at speed $v_\textrm{jet}$.  This is indistinguishable from the situation considered by \citet{1967MNRAS.135..345R}.
Our conclusion for this section is that the data are consistent with the KO model. It should be clear from Fig.~\ref{plt:cartoon} that the phase coherence we observe in PKS~2131$-$021 across a wide range of observing frequencies is due to the fact that $\theta \ll 1$.  Were this not the case, the light travel time arguments from the emission regions to the observer illustrated in Fig.~\ref{plt:cartoon} would not apply and coherence between different frequency bands would be lost.

An important question is ``At what level does one expect to see harmonics of the fundamental frequency?''.  This is beyond the scope of the present paper, and  will be discussed in a future paper,  For the present, we refer the reader to Appendix \ref{sec:higher2}.

An interesting physical issue is why the dominant frequency emitted along the jet should depend exponentially on $r'$.  The requirement of a well-defined spectral peak at each $r'$  (as well as compactness and flat spectrum of the emitting region) suggests a transition from optically thick to thin radiation due to synchrotron self-absorption.  It is hard to accomplish this if the jet properties are insensitive to radius, since the transition will not be abrupt enough, leading to phase mixing (e.g., resulting in strong 2.7~GHz radiation from a large range of radii) which will damp out the amplitude of the sinusoid.  The most likely jet property to vary downstream is the magnetic field $B$ in the jet and, in order to get the high frequency radiation to lead the low frequency in phase, we need the magnetic field to decrease exponentially along the jet.

This type of model offers  an interesting working hypothesis. For example, one could imagine that some event (e.g., a recollimation shock or onset of Kelvin--Helmholtz instability) stirs up turbulence and turbulent resistivity, allowing the current flowing down the jet to leak out sideways.  If this happens exponentially with $r'$, it will lead to an exponential decrease in $B$ and a corresponding exponential decrease in $\nu$ with $r'$.  This is an interesting physical problem in its own right, since it is usually assumed that the current has to leak out of the jet somewhere in order to close the circuit---this would suggest that the leakage could occur very close to the central engine.

\section{What Fraction of Blazars are SMBHB Candidates?}\label{sec:fraction}

\begin{figure}[!t]
   \centering
   \includegraphics[width=1.0\linewidth]{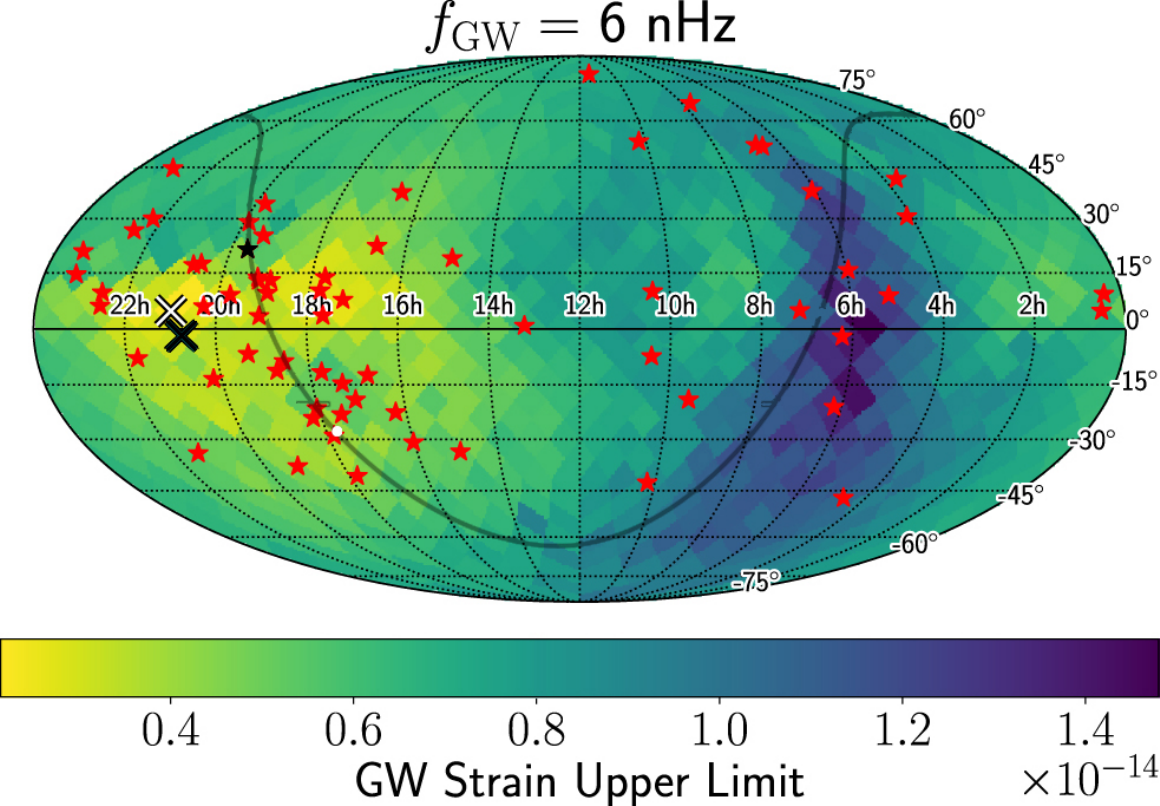}
      \caption{Mollweide projection of the sky in right ascension and declination coordinates, showing the position of 
      PKS~2131$-$021 (black cross) plotted on the NANOGrav strain upper limits \citep{2023ApJ...951L..50A}.   The gray curve shows the Galactic Plane. The white dot marks the location of the Galactic center.  The  white cross marks the position of the highest NANOGrav sensitivity.    The red asterisks mark the positions of the millisecond pulsars used in the \citet{2023ApJ...951L..50A} analysis, from which this figure is adapted. The black asterisk marks the position of 4C 21.53, the first millisecond pulsar (see text).
      } 
         \label{plt:nanograv}
\end{figure}

The recent discovery of a stochastic background of gravitational waves (GW) with periods of months to years \citep{2023ApJ...951L...8A,2023A&A...678A..50E} using millisecond pulsars  as GW probes has promoted an intense search for the GW electromagnetic counterparts \citep[e.g.,][]{2023ApJ...951L..50A}, which makes clear the importance of SMBHB searches such as the one we are engaged in with the OVRO 40 m Telescope.  Millisecond pulsars \citep{1974MmRAS..78....1R,1982Natur.300..615B}\footnote{There were two crucial steps in the discovery of millisecond pulsars: (i) the discovery of interplanetary scintillation, at Galactic latitude $-0.3^\circ$, in 4C 21.53 by \citet{1974MmRAS..78....1R} (see \citet{2024JAHH...27..453R}),  which first drew attention to the singular nature of this object; and (ii) the discovery of millisecond pulses from 4C 21.53W by \citet{1982Natur.300..615B}.} enable searches for gravitational waves with periods of months to years using pulsar timing arrays such as the European pulsar timing array \citep[\hbox{EPTA},][]{2023arXiv230616226A}, the North American Nanohertz Observatory for Gravitational Waves  \citep[NANOGrav,][]{2023ApJ...951L...8A,2023ApJ...952L..37A,2023ApJ...951L..50A},  the Parkes pulsar timing array \citep[\hbox{PPTA},][]{2023PASA...40...49Z}, and the MeerKAT Pulsar Timing Array \citep[\hbox{MPTA},][]{2025MNRAS.536.1489M}. As it happens, PKS~2131$-$021 is almost optimally located for study with the \hbox{EPTA}, NANOGrav, and the \hbox{PPTA}.  In Fig.~\ref{plt:nanograv}, the black and white crosses mark PKS~2131$-$021 and the location of greatest NANOGrav sensitivity, respectively \citep{2023ApJ...951L..50A}. 

It is clearly important to determine the fraction of blazars that are strong SMBHB candidates. \citet{Sandrinelli2017,2018AandA...615A.118S} estimated that $\sim 10\%$ of $\gamma$-ray bright blazars are QPO candidates, but they later  determined that all of their candidates were likely produced by the red noise tail \citep{2019MNRAS.482.1270C}.  \citet{Ackermann2015} reported QPOs in PG 1553+113 based on the Fermi-LAT $\gamma$-ray light curve over a period of 6.9 years.  This periodicity has recently been confirmed using 15 years of Fermi-LAT data \citep{2024ApJ...976..203A}, see also \citet{2024A&A...686A.300A}. In a multi-wavelength study \citet{2024MNRAS.529.3894M} used the above GLS analysis (with 20,000 simulations) to show that the QPOs in PG 1553+113 have a global $p$-value $=1\times 10^{-3}\;(3.2 \sigma)$ in the Fermi light curve, and  $p$-value $=3.4\times 10^{-2}$ in the OVRO 15 GHz light curve. Thus PG 1553+113 passes the $3\sigma$ global $p$-value test at $\gamma$-ray energies and is a strong SMBHB candidate, according to the criterion we adopted in Paper 1.

\citet{2024arXiv240802645D} have similarly shown that PKS J0805$-$0111 is a strong SMBHB candidate, based on the OVRO 15 GHz light curve. They also found  hints of the same periodicity in the Fermi-LAT light curve. 

Thus, in the complete statistical sample of 1158 blazars being monitored at OVRO \citep{Richards2011}, three strong SMBHB candidates have been identified thus far: one of them (PG 1553+113) exhibits clear periodic oscillations at $\gamma$-ray energies and hints of periodic oscillations at radio energies; while the other two (PKS J0805$-$0111 and PKS~2131$-$021) exhibit clear periodic oscillations at radio energies and hints of periodic oscillations at $\gamma$-ray energies.  It is clear, therefore, that the fraction of blazars that are strong SMBHB candidates, with periods in the range of a few months to $\sim$ five years, is  $\gtrsim 1$ in 400.

We are in the process of carrying out a detailed study applying the above GLS analysis to all of the $\sim 1800$ AGN in the 40 m monitoring sample. Since this study is in progress we give here only a brief outline of our findings thus far. We have completed the local significance analysis on about half of the objects in the sample, and we find that about 15\% of these pass the $3\sigma$ local  significance  threshold.  We have yet to  carry out the global significance analysis on these objects. In the case of PKS~2131$-$021 the ratio of the local $p$-value to the global $p$-value is $>$40.  If we assume that the same applies to the other objects in the sample then we project that at most 20 of the objects in our sample will pass the   $3\sigma$ global significance threshold.  We have thus far found four AGN that pass the $3\sigma$ global significance threshold.  Thus the projected fraction of AGN in our sample that are strong SMBHB candidates, based on the radio data alone, is at least  1 in 500. 

Thus far we have picked out only the strongest SMBHB candidates - i.e. those that show very clear periodicity at $\gamma$-ray or radio frequencies.  Clearly, there must be other SMBHB candidates in which the periodicities do not stand out so clearly against the background of strong variability we see in all blazars. It should be borne in mind that the test we are applying to the radio light curves is selecting SMBHB candidates in which the sinusoidal variation is both strong and persists over most of the 16 year period of observations, from 2008 to 2023, inclusive.  We have seen in PKS~2131$-$021 that the sinusoidal variation was absent for 19 of the past 48 years. In addition the sinusoidal variations have to be strong enough,  compared to the random variations in the source, to yield a $3\sigma$ global significance. It appears very likely that there will be AGN that are SMBHBs in which there are sinusodial signals that do not pass this threshold.  For all of these reasons, in our view, the fraction of SMBHBs among jetted-AGN is likely to be $\sim 1$ in 100, or higher. 

Such estimates from EM surveys can be compared to the observed value of the amplitude of the GW background, under the assumption that its origin is an ensemble of SMBHs \citep[e.g.,][]{2018MNRAS.481L..74H,2018ApJ...863L..36I,2020ApJ...900L..42N,2018ApJ...856...42S}.
Based on an assumed quasar mass function and SMBHB candidates from the Catalina Real-Time Transient Survey, \citet{2024arXiv240519406C} estimated an upper limit of $25\%$ on the fraction of quasars hosting an SMBHB.
\cite{2018MNRAS.481L..74H} presented an upper limit on the fraction of blazars that host an SMBHB with a period less than $5\,$yr assuming certain blazar population properties.
Comparing this to an older upper limit on the GW amplitude at a frequency $f=1\,$yr of $A_{\rm{yr}}<1.45\times10^{-15}$ at the 95\% credible level \citep{2018ApJ...859...47A} from the 11yr of NANOGrav dataset, they estimated a binary fraction of $\lesssim0.1\%$.
However, the updated GW amplitude values based on the NANOGrav 15yr dataset of $A_{\rm{yr}}=2.4^{+0.7}_{-0.6}\times10^{-15}$ at the 90\% level \citep{2023ApJ...951L...8A} implies a revised  binary fraction to $\lesssim 0.5\%$.
These estimates are roughly consistent with our estimate above.

It is interesting to note that JWST is finding a large number of dual AGN at high redshifts, some of which are separated by a few or several hundred parsecs
\citep{Perna2023,Marshall2023,Ubler2024,Maiolino2023}.
The fraction of these dual AGN seems to be fairly high (20--30\%), while the majority of AGN hosts have close companions or show post-merging signatures.
While the fate of black holes in interacting galaxies at large (several-kiloparsec) separations is not yet clear \citep{DiMatteo2023}, black holes with separations of less than 1\,kpc are expected to become black hole binaries on sinking times of a few to several hundred million years, especially given the high central stellar densities of these early galaxies, which increase the drag force on BHs. A fraction of such binary BHs with appropriate jet emission and orientation should be detectable as blazars, with periodic signals similar to those seen in PKS~2131$-$021 at $z=1.283$. It is difficult to compare the fraction of blazars that are SMBHB candidates and the fraction of candidate BH mergers from recent JWST observations with theoretical models and simulations of BH merging, due to poor statistics and large uncertainties in both  observations and simulations. However, these are all areas of active and rapid development, and a quantitative assessment might be possible in the near future.

\section{Conclusion and Future Work}\label{sec:discussion}

The most important results of this paper are
\begin{enumerate}
\item The demonstration of continuing coherent sinusoidal variations in the flux density of  PKS~2131$-$021 for a further 2~years.

\item An increase in the significance of the periodicity in PKS~2131$-$021 by a factor of 3.

\item The discovery that coherent strong sinusoidal variations in the flux density of PKS~2131$-$021 extend from below 2.7\,GHz to optical wavelengths.

\item The discovery that the phases of the sinusoidal variations show a monotonically increasing lag with increasing observing wavelength. 

\item The discovery that the short-term random fluctuations at optical wavelengths are an order of magnitude or more greater than those at radio wavelengths, when compared with the amplitudes of the sinusoidal signals in these two wavebands.

\item The data are consistent with the KO model, in which the sinusoidal variation is caused by aberration due to the orbital motion of a jet-producing SMBH in an \hbox{SMBHB}.

\item Given the critical importance of the identification  of the origin of the stochastic background of gravitational waves detected by PTAs, only carefully-vetted strong SMBHB candidates, of proven high statistical significance, should be used in  PTA targeted searches.

\item The periodicity at radio wavelengths has remained persistent and stable across several epochs. The newly discovered periodicity in optical and possibly in $\gamma$-rays again shows the robustness of this discovery. We thus predict that this periodicity will continue to be observed in this object over a wide range of frequencies.

\end{enumerate}

It is clear that the periodicity we see in PKS~2131$-$021 is telling us something fundamental about this object.  The possible causes of this periodicity that should be considered are
(i)~orbital motion of an \hbox{SMBHB};
(ii)~precessional motion of the accretion disk; and 
(iii)~a magneto-hydrodynamic (MHD)  instability.
We believe that (iii) can be ruled out by the fact that the sinusoidal variations disappeared for~19~years and then re-appeared with the same period and in phase with the original variations (see Paper~1).  This is evidence of a ``clock'' continuously ticking during the 19-year gap in the sinusoidal variations.  Of the remaining options (i) and (ii), neither can be ruled out, but it is clear that orbiting motion in an SMBHB is a distinct possibility. For this reason we believe that pulsar timing arrays should carry out coherent searches for gravitational waves from PKS~2131$-$021 and other strong well-vetted OVRO SMBHB candidates.

The Atacama Cosmology Telescope (ACT) \citep{2023ApJ...956...36L}, and the South Pole Telescope (SPT)\citep{2021ApJ...916...98G} have  accumulated many years of data on  tens of hundreds of blazars, which provide invaluable information on SMBHB candidates, and may well reveal new SMBHB candidates as these data are mined. With  large-scale sky surveys, such as with the ZTF, the Vera Rubin Observatory, and the Simons Observatory \citep{2019JCAP...02..056A}, hundreds of thousands of blazars can be monitored for long periods. We now have several other strong SMBHB candidates in the  OVRO monitoring sample, and, if the KO model  is correct, there must be several SMBHB candidates that are difficult to detect because of the dependence of the fractional variation in flux density on the angle of inclination of the orbital plane to the line of sight.   Thus we estimate that the fraction of radio blazars that are SMBHB candidates is at least 1 in 100, which  raises the clear possibility of detecting thousands of SMBHB candidates with well-determined periods. 

An important next step in exploring sources like PKS~2131$-$021 is to seek spectroscopic manifestation of the periodicity. This is an approach which has been highly productive in many Galactic binary sources, most notably SS~433 \citep[see, e.g.,][]{2006Ap&SS.304..271S} which exhibits ``stationary'' and ``moving'' spectral lines plus strong evidence for a third body, all of which exhibit many periodic behaviors reflecting an underlying, dynamical clock. 

There are many ways in which variable emission lines might arise in sources such as PKS~2131$-$021. These include launching a high speed wind from an orbiting accretion disk, either radiatively or magnetically; forming lines in an expanding and cooling high-speed jet which might orbit and/or precess and or be perturbed by the companion black hole; and having the proposed black hole binary being accompanied by stars or even a third black hole. Indeed, it would be rather surprising if the various dynamical explanations, that we have entertained to account for the periodicity in the flux, produced no associated spectroscopic variation. 

We are aware of three, relatively low resolution, optical spectra of PKS~2131$-$021 exhibiting prominent broad lines of  \ion{C}{3}],  \ion{C}{2} and \ion{Mg}{2}, taken with different telescopes. There do appear to be changes, but it is hard to determine if these are significant. What is needed is a small number of relatively high dispersion, $R\sim5000$--$10000$, and high signal-to-noise ratio spectra taken with the same instrument. It is not necessary to observe for a full 4.8 yr period. Instead, we should seek changes over months which can be related to the observed period and phase of the flux density measurements. In addition to more optical spectroscopy, near infrared observations, which have lower absorption and where Balmer and Paschen lines should be observable, would be instructive.

If the Kinetic Orbital (KO) model is correct, then the optical sinusoidal curves will give us the expected phase of the GW waves from the SMBHB candidates.  At this point only the expected amplitude of the GW will be unknown.  Clearly, were a method to be developed for measuring the speeds of the SMBHBs, then a fully-coherent search for GW from the combination of ${\sim}100$ SMBHB candidates could be mounted in which the period, the phase, and the amplitude of the expected GW signal from each candidate could be added coherently.  In such circumstances, the chances of being able to confirm, the SMBHB origin of the stochastic background of GW would be greatly increased.


\section*{ACKNOWLEDGEMENTS}
This work is supported by NSF grants AST2407603 and AST2407604. We thank the California Institute of Technology and the Max Planck Institute for Radio Astronomy for supporting the  OVRO 40\,m program under extremely difficult circumstances over the last 8~years in the absence of agency funding. Without this private support these observations could not have been made. We thank Adam Hincks for useful discussions. We also thank all the volunteers who have enabled this work to be carried out. 
Prior to~2016, the OVRO program was supported by NASA grants \hbox{NNG06GG1G}, \hbox{NNX08AW31G}, \hbox{NNX11A043G}, and \hbox{NNX13AQ89G} from~2006 to~2016 and NSF grants AST-0808050 and AST-1109911 from~2008 to~2014.
S.K. \& K.T. acknowledge support from the European Research Council (ERC) under the European Unions Horizon 2020 research and innovation programme under grant agreement No.~771282.
I.L and S.K. were funded by the European Union ERC-2022-STG - BOOTES - 101076343. Views and opinions expressed are however those of the author(s) only and do not necessarily reflect those of the European Union or the European Research Council Executive Agency. Neither the European Union nor the granting authority can be held responsible for them.
W.M. acknowledges support from ANID project Basal FB210003. A.S.  and R.B acknowledge support by a grant from the Simons Foundation (00001470,RB,AS). R.R. and B.M. and P.V.d.l.P. acknowledge support from ANID Basal AFB-170002, from Núcleo Milenio TITANs (NCN2023\_002), and CATA BASAL FB210003. P.V.d.l.P. also acknowledges support by the National Agency for Research and Development
(ANID) / Scholarship Program / Doctorado Nacional/2023--21232103.
C.O. acknowledges support from the Natural Sciences and Engineering Research Council (NSERC) of Canada.
T.H. acknowledges support from the Academy of Finland projects 317383, 320085, 322535, and 345899.
The NANOGrav collaboration receives support from National Science Foundation (NSF) Physics Frontiers Center award numbers 1430284 and 2020265.
Part of this research was carried out at the Jet Propulsion Laboratory, California Institute of Technology, under a contract with the National Aeronautics and Space Administration. 
This paper makes use of the following ALMA data: ADS/JAO.ALMA\#2011.0.00001.CAL. ALMA is a partnership of ESO (representing its member states), NSF (USA) and NINS (Japan), together with NRC (Canada), NSTC and ASIAA (Taiwan), and KASI (Republic of Korea), in cooperation with the Republic of Chile. The Joint ALMA Observatory is operated by ESO, AUI/NRAO and NAOJ. The National Radio Astronomy Observatory is a facility of the National Science Foundation operated under cooperative agreement by Associated Universities, Inc. This research has made use of the NASA/IPAC Extragalactic Database (NED), which is funded by the National Aeronautics and Space Administration and operated by the California Institute of Technology.

\appendix

\section{Sine-wave fits to the optical data}\label{sec:sineoptical}

We fitted the following sine-wave model to the optical data:
\begin{equation}
m(t) = A \sin(\phi-\phi_0) + m_0,
\end{equation}
where $\phi = 2\pi (t-t_0)/ P$, and $t_0 = 59000$. The best-fit parameters were found by maximizing the log-likelihood function defined as
\begin{equation}
\ln\mathcal{L}=-\frac{1}{2}\sum_{g,r,i}\sum_{j=1}^{N_{\rm obs}}\left(\frac{(m_j-m_j^{\rm model})^2}{\sigma_j^2+\sigma_0^2}+\ln(\sigma_j^2+\sigma_0^2)\right).
\end{equation}
First, we assumed that the period and phase are identical for all filters, but the amplitudes, mean magnitudes, and $\sigma_0$ may depend on the passband (model~1). Subsequently, we relaxed this assumption and allowed the sine phases to depend on the filter (model~2). Results of the fits are reported in Table~\ref{tab:sine_wave_optical}.

\begin{table}[ht]
\centering
\footnotesize
\caption{Results of the sine-wave fits to the optical data}
\label{tab:sine_wave_optical}
\begin{tabular}{lrr}
\hline \hline
Parameter & Value \\
\hline
$P$ (d)            & $1764 \pm 36$      & $1762 \pm 36$ \\
$\phi_0$ (rad)     & $4.372 \pm 0.056$  & \dots \\
$\phi_0$ ($r$)       & \dots              & $4.353 \pm 0.080$ \\
$\phi_0$ ($g$)       & \dots              & $4.349 \pm 0.084$ \\
$\phi_0$ ($i$)       & \dots              & $4.456 \pm 0.113$ \\
$m_0$ ($r$)          & $17.454 \pm 0.021$ & $17.453 \pm 0.021$ \\
$A$   ($r$)          & $0.358 \pm 0.032$  & $0.358 \pm 0.032$ \\
$\sigma_0$ ($r$)     & $0.363 \pm 0.015$  & $0.363 \pm 0.015$ \\
$m_0$ ($g$)          & $18.030 \pm 0.022$ & $18.030 \pm 0.022$ \\
$A$   ($g$)          & $0.366 \pm 0.033$  & $0.366 \pm 0.034$ \\
$\sigma_0$ ($g$)     & $0.383 \pm 0.016$  & $0.383 \pm 0.016$ \\
$m_0$ ($i$)          & $17.146 \pm 0.035$ & $17.155 \pm 0.037$ \\
$A$   ($i$)          & $0.379 \pm 0.055$  & $0.388 \pm 0.058$\\
$\sigma_0$ ($i$)     & $0.376 \pm 0.023$ & $0.376 \pm 0.023$\\
\hline
$\chi^2/\mathrm{d.o.f.}$ & $785.4/780$ & $786.7/778$ \\
\hline
\end{tabular}
\end{table}

\begin{figure}
\includegraphics[width=1.0\textwidth]{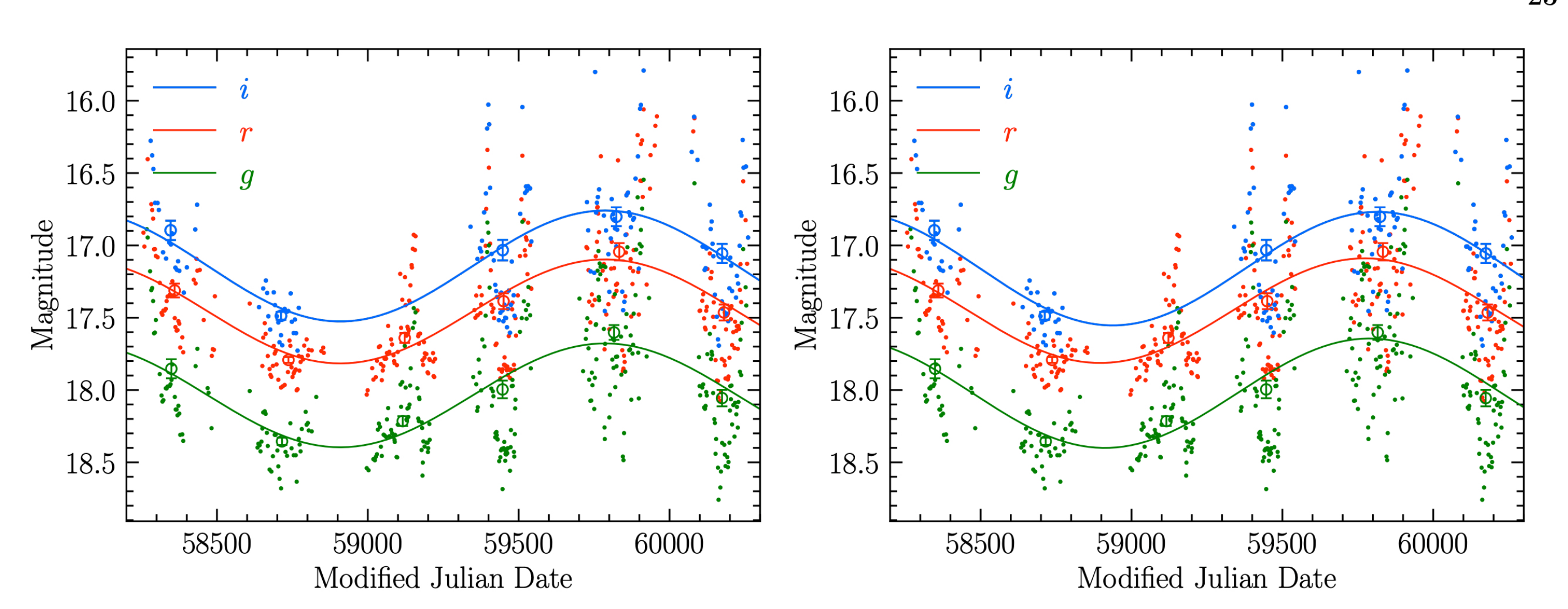}
\caption{Best-fit sine-wave models to the optical data, assuming that the periods and phases are identical in all filters (left panel) or allowing the phases to depend on the passband (right panel).}
\end{figure}

\section{Fit of sinusoidal model using simulation-based likelihood}\label{sec:correlated}
As shown in Fig.~\ref{plt:lightcurves2}, the OVRO and Haystack
15 GHz lightcurves appear to fit the same sine wave, albeit with
different amplitudes and zero points. We can model these as
\begin{align}
S_i^O &= A^O\sin(2\pi[(t_i^O - t_0)/P]+\phi) + S_0^O + n_i^O \notag \\
S_i^H &= A^H\sin(2\pi[(t_i^H - t_0)/P]+\phi) + S_0^H + n_i^H \label{eq:data-model}
\end{align}
Where $i$ is a sample index (different range for OVRO and Haystack),
$t_i$ is is the observation time of sample $i$, and $n_i$ its noise,
and the superscripts $O$ and $H$ refer to OVRO and Haystack respectively.
The unknown parameters are the period $P$, phase $\phi$, amplitudes
$A^O$ and $A^H$ and the flux density offsets $S_0^O$ and $S_0^H$.

\subsection{The noise covariance matrix}\label{sec:covariance}

\begin{figure}[htp]
\begin{minipage}[c]{0.45\linewidth}
\centering
\scalebox{1}[-1]{\includegraphics[width=0.94\linewidth]{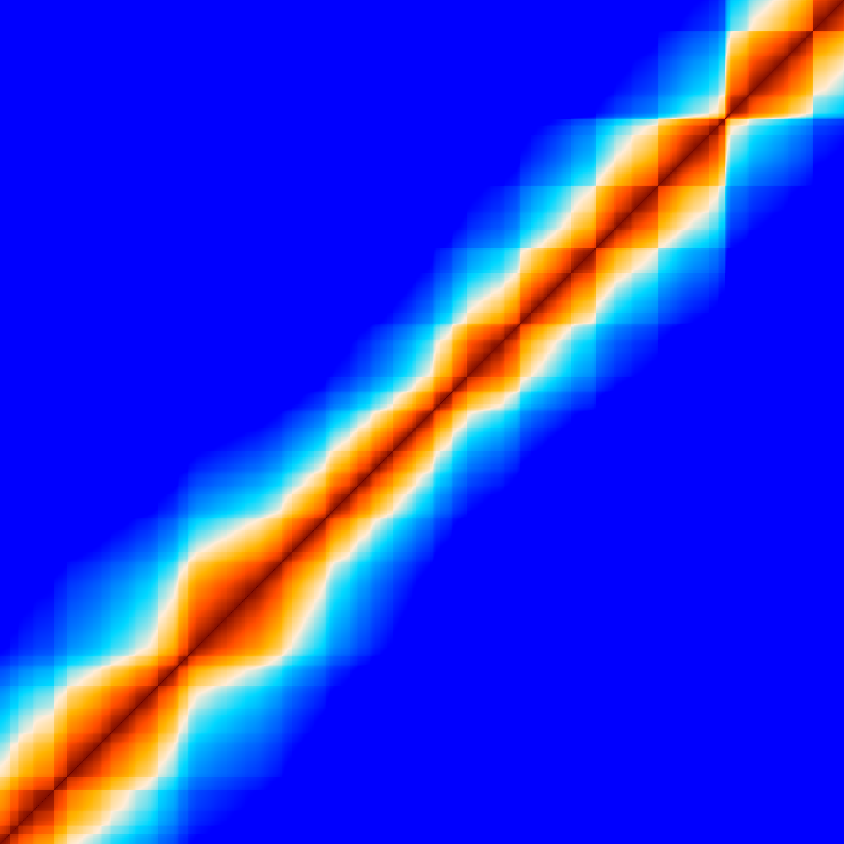}}
\caption{Correlation part of the sample-sample covariance matrix $N^O$ for the OVRO lightcurve. Dark blue is zero, dark red is one. The blocky structure is caused by the uneven sample spacing.}
\label{fig:nmat}
\end{minipage}
\hfill
\begin{minipage}[c]{0.45\linewidth}
\includegraphics[width=\linewidth]{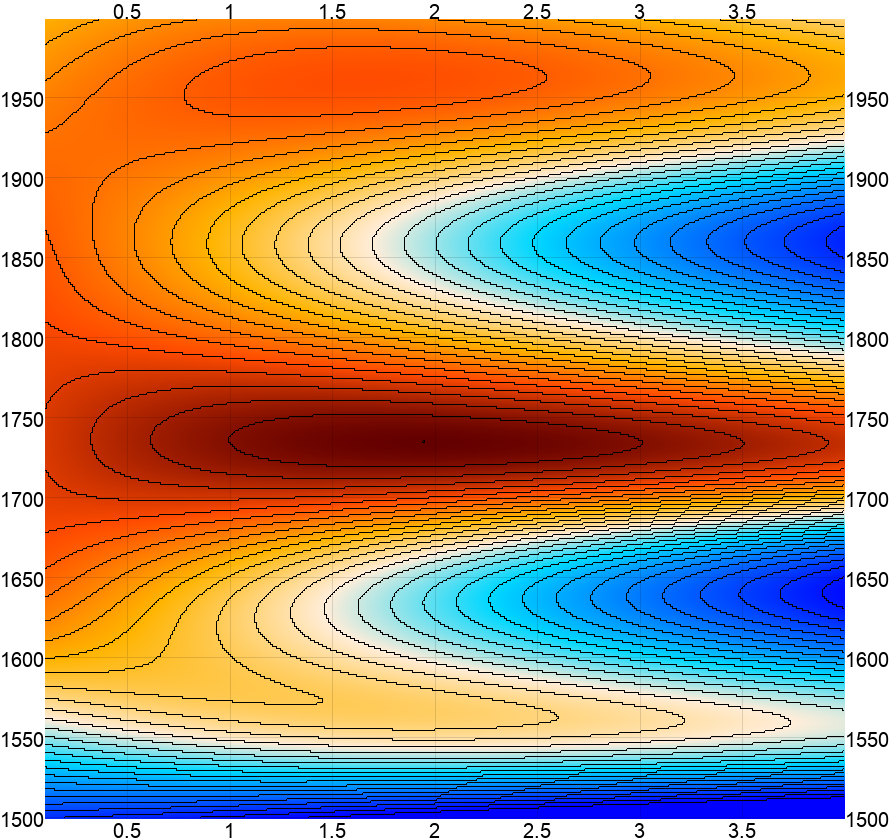}
\caption{$\log P$ plotted as a function of $A^{HO}$ (horizontal axis, the Haystack amplitude
relative to OVRO) and $P$ (vertical axis, the sine period in days). The black contours are steps of 1 in $\log P$.}
\label{fig:posterior-grid}
\end{minipage}
\end{figure}

The noise includes the instrument and atmospheric noise and any AGN variability that is not part of the sinusoidal variation. We approximate the noise as Gaussian, with covariances $N^O$ and $N^H$. These covariance matrices are challenging to estimate due to the limited amount of data available, but we can use as a proxy the lightcurve simulations. Hence we build the sample-sample covariance matrixes $N^O_{ij}$ and $N^H_{ij}$. These covariance matrices originally had some unphysical properties caused by limitations in the simulation methodology: As the point separation grows, the correlation falls as expected, but eventually becomes negative before starting to rise again. The rise is caused by the periodic boundary conditions implied by the Fourier methods used when constructing the simulations, while we interpret the negative correlation region as being caused by the characteristic length scale of the sinusoidal signal itself. Neither of these properties is relevant to the AGN red noise we wish to model, so we modified the covariance matrices to set the negative and rising areas equal to zero.\footnote{We have confirmed that the unphysical periodicity of the simulations did not impact the significance estimate the simulations were originally built for.} The resulting OVRO covariance matrix is shown in Fig.~\ref{fig:nmat}.

\subsection{Likelihood}
The model in equation~\ref{eq:data-model} is inefficient to fit because it has four
non-linear parameters: $P$, $\phi$, $A^O$ and $A^H$. We can reduce this
to two by re-parameterizing as follows:
\begin{align}
S_i^O &= \alpha\sin(\theta_i) + \beta\cos(\theta_i) + S_0^O + n_i^O \notag \\
S_i^H &= A^{HO}\Big(\alpha\sin(\theta_i) + \beta\cos(\theta_i)\Big) + S_0^H + n_i^H
\end{align}
where $\theta_i = 2\pi(t_i - t_0)/P$, and $A^{HO} = A^H/A^O$ is the ratio of the Haystack
amplitude to the OVRO amplitude. With this parameterization the only non-linear
parameters are $P$ and $A^{HO}$. Given these parameters, the rest can be
evaluated in a single, linear step. We can write the model in linear algebra form
\begin{align}
S &= Qa + n
\end{align}
Here $S$ and $n$ are stacks of the OVRO and Haystack data and noise respectively; $a = [\alpha,\beta,S_0^O,S_0^H]$
are the set of linear parameters; and $Q$ is a matrix that depends on $P$ and $A^{HO}$ and
implements equations B3 and  B4. We can then write the Likelihood as
\begin{align}
-2\log L &= |2\pi N| + (S-Qa)^TN^{-1}(S-Qa) \notag \\
&= |2\pi N| + (a-\hat a)^T A^{-1} (a-\hat a) + S^TN^{-1}S - \hat a A^{-1} \hat a
\end{align}
where $\hat a = (Q^TN^{-1}Q)^{-1}Q^TN^{-1}S$ and $\hat A^{-1} = Q^TN^{-1}Q$.
These parameters have simple interpretations:
$\hat a$ is the maximum-likelihood estimator for the linear parameters, and $\hat A$
is the covariance of this estimator. To reduce the non-linear fit to just two parameters
we must marginalize over the linear parameters, which means integrating over all values
of $a$. Since $a$ has a Gaussian likelihood with covariance $\hat A$, we know that its
integral is  $\sqrt{|2\pi \hat A|}$, so the marginalized likelihood is
\begin{align}
-2\log L' &= |2\pi N| - |2\pi \hat A| + S^TN^{-1}S - \hat a^T \hat A^{-1} \hat a
\end{align}
The first and third terms don't depend on any of our parameters and so can be
ignored in an MCMC fit. Additionally ignoring the second term is equivalent to
a Jeffrey's minimum-information
prior on $a$. We are therefore left with the following posterior probability for
the non-linear parameters.
\begin{align}
\log P &= \frac12 \hat a^T \hat A^{-1} \hat a + \log \textrm{prior} \label{eq:logP}
\end{align}
where ``prior'' is a manual prior on the parameters. We used $P\in [1000,3000]$ and $A^{HO}\in [0.1,4]$
in our fit.

The $P,A^{HO}$ posterior is shown in Fig.~\ref{fig:posterior-grid}.

\subsection{Full MCMC search}
\begin{figure}[htp]
\centering
\includegraphics[width=\textwidth]{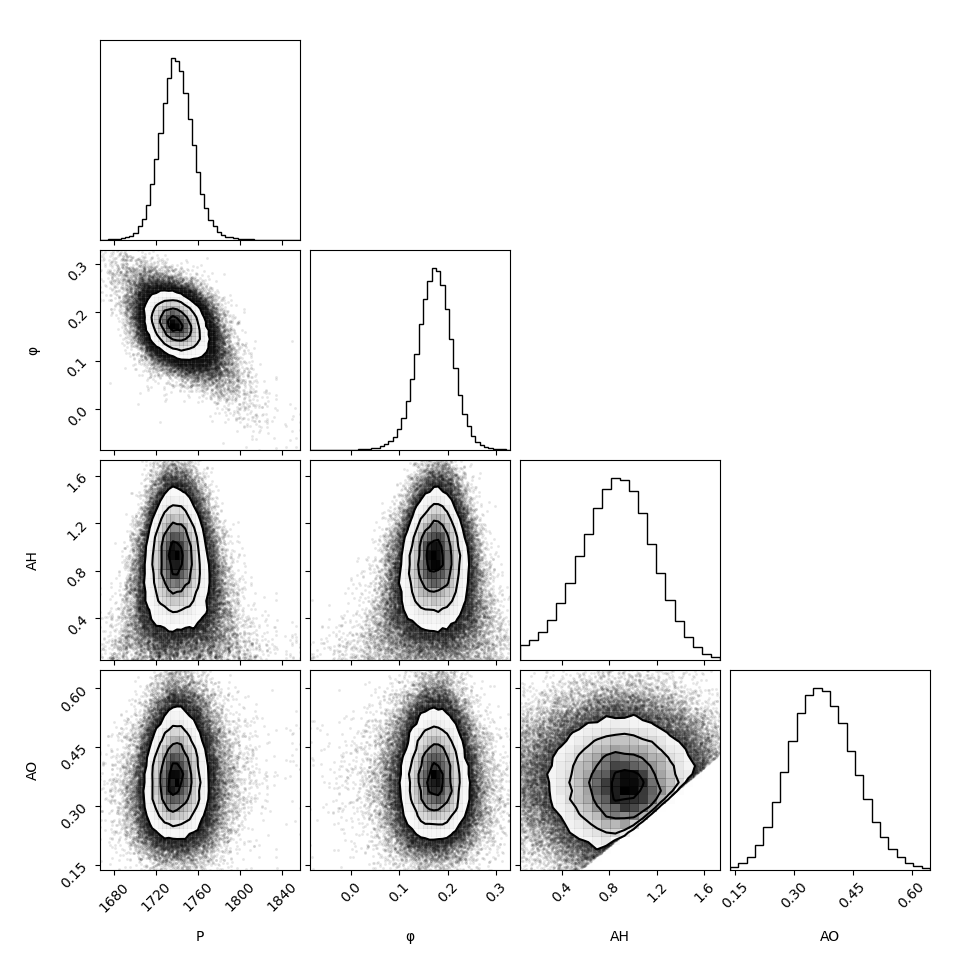}
\caption{Distribution of samples from MCMC fit of the parameters in equation~\ref{eq:data-model}. Note: Unlike elsewhere in the text, $\phi$ is here reported in units of periods instead of radians, so it's smaller by a factor $2\pi$.}
\label{fig:corner}
\end{figure}

We performed a full MCMC search of the posterior in equation~\ref{eq:logP} using \emph{emcee} with 4 walkers, a burn-in of 30\,000 samples and 30\,000 output samples. For each non-linear parameter sampled this way, we drew a sample of the linear parameters from their conditional distribution. Fig.~\ref{fig:corner} shows a corner plot for the resulting samples. Our final fit parameters are $P = 1739.8 \pm 17.4$ days, $\phi = 1.078 \pm 0.025$ radians offset from MJD $t_0 = 51000$, $A^O = 0.374\pm0.083$ Jy, $A^H = 0.853\pm 0.304$.

This fit is quite sensitive to the details of the noise covariance matrix. This matrix could be improved by re-estimating the noise properties after subtracting the best-fit signal model. As it is, the noise matrix is somewhat contaminated by the sinusoidal signal. We leave this for a future improvement, and for now note that the error bars are likely to be overestimated.

\clearpage

\section{KO model Parameters for the Sinusoidally Varying Core Component}
\label{sec:brightness}
\citet{2021ApJ...923...67H} were specifically interested in the PKS~2131$-$021 component brightness temperatures and  they fitted all of the core regions  in a consistent fashion, taking particular care not to mix components between epochs.  The details of their approach are given in the methods section  of their paper.   They found a median $T_{\rm b,\,observed} =  5.8 \times 10^{11} $K over 24 epochs from 1995 to 2013.  They also found an estimated intrinsic median brightness temperature for their sample of $T_{\rm b,\, intrinsic} = 4.1 \times 10^{10} $K, which gives an estimated Doppler factor of $D=T_{\rm b,\,observed}/T_{\rm b,\, intrinsic}=14$.  Thus, using the value $\theta=3.8^\circ$ determined by \citet{2021ApJ...923...67H}, for this paper we adopt the value $\beta=0.995$, corresponding to $\gamma=10$ and $D=14$.     The amplitude of the 15 GHz OVRO light curve, assuming a fixed period $P=1739.2$ days for the sinusoidal component, is 0.42 Jy.  In order not to detect it, we need to suppress the amplitude of the first harmonic to below $0.1 \times$ the amplitude of the fundamental.   Hence the amplitude of the first harmonic should be less than 42 mJy. A sinusoidal first harmonic of amplitude 42 mJy is shown in Fig.~\ref{plt:harmonic}. To achieve this suppression of the amplitude of the first harmonic, we need a value of $\beta_1 \ll 1$, so we adopt a value $\beta_1=0.020$.  

Note that from the MOJAVE results \citep{2021ApJ...923...67H}, which find an apparent speed of the fastest moving components of $19.9$c$\pm 2.6$c, we might expect to use a value of $\gamma\sim 20$. However, the  structure that is making the sinusoidal variations is inside the core component.  We have no guarantee that the $\gamma$ value of the components within the core is the same as that of those outside of the core.  On the KO model, a $\gamma$ of 20 would make the radiation emission regions 4 times further away from the SMBH than a $\gamma$ of 10, and we would like to try to keep this distance as small as allowed by the observations.  The smallest value of $\gamma$ that gives the observed fractional variation is $\gamma=10$, so we adopt this value.

We now determine the equipartion angular size and the equipartition brightness temperature following \citet{1977MNRAS.180..539S} and \citet{1994ApJ...426...51R}, and using the updated expressions for the ${\rm \Lambda}$CDM cosmology given by \citet[][Appendix B]{2021ApJ...907...61R}. In the course of this work we realized that there is a typographical error in the expression for $F(\alpha)$ given in \citet{1977MNRAS.180..539S}.  The correct expression, using the $\alpha$ definition $S\propto \nu^{- \alpha}$ of \citet{1977MNRAS.180..539S}, is
$F(\alpha)= 1.6 C(\alpha) [(\nu_{\rm low}^{1/2-\alpha}- \nu_{\rm high}^{1/2-\alpha})/((2\alpha -1)f_1(\alpha))]^{1/17} f_3(\alpha)^{-7/17}  $, where $\nu_{\rm low}$  and $\nu_{\rm high}$ are the lower and upper cutoffs in the spectrum. The typographical error in \citet{1977MNRAS.180..539S} is that the exponent $(-7/17)$ that applies to  $f_3(\alpha)  $ was omitted.   Values of $C(\alpha)$ are tabulated by \citet{1977MNRAS.180..539S}, and the expression for $f_3(\alpha)$ is given by equation 15c of \citet{1968ARA&A...6..321S}, and plotted by \citet[][Appendix B]{2021ApJ...907...61R}, but note the change in the sign of $\alpha$ between these two papers. We have recalculated the expressions for the equipartition angular size, $\psi_{\rm eq}$, and equipartition brightness temperature, $T_{\rm eq}$, to higher accuracy than  given in \citet{2021ApJ...907...61R}, and in so doing discovered an error of $2^{1/17} = 1.042$.  The corrected/revised expressions are given below.

In the ${\rm \Lambda}$CDM cosmology the equipartition angular size becomes:
\begin{equation}\label{psi1}
\psi_{_{\rm eq}} = 1.74\,   r^{-{1 \over17}} S^{8 \over 17}  \nu^{-{{35+ 2\alpha} \over 34}}   (1+z)^{{15-2\alpha} \over 34}
 F(\alpha) \, ,
\end{equation}
where $r$ is the comoving coordinate distance in gigaparsec, $S$ is the flux density in jansky at the peak of the spectrum,  $\nu$ is the frequency at the peak in the spectrum in MHz,  and $F(\alpha)$ is given in \citet{1977MNRAS.180..539S}, noting that they use the convention $S \propto \nu^{-\alpha}$.  Hence we find $\psi_{_{\rm eq}}=0.65$ mas for the core of PKS~2131$-$021.

The equipartition brightness temperature is given by 
\begin{equation}\label{Teq}
T_{\rm eq} = 5.84\times10^{11} \biggl[{{r\over {(1+z)}}}\biggr]^{2/17} F(\alpha)^{-2} 
 (1+z)^{(2\alpha - 
13)/17}S^{1/17}\bigl(10^3\nu\bigr)^{(1+2\alpha)/ 
17}{\rm 
K} \, .
\end{equation}
and hence we find $T_{\rm eq} = 1.98\times 10^{10}$\,K, Note that this is a factor 2 lower than the value for $T_{\rm b,\, intrinsic}$ adopted above.  However the brightness temperatures that \citet{2021ApJ...923...67H} calculate are the Gaussian brightness temperatures, which must be reduced by a factor 1.8 to compare with those of  \citet{1994ApJ...426...51R}. This reduces the $T_{\rm b,\, intrinsic}$ to $2.3 \times 10^{10}$ K,   which implies a ratio of particle to magnetic field energy densities of $\sim 4$ \citep{1994ApJ...426...51R}. 

\section{The Absence of Higher Harmonics}\label{sec:higher2}

\begin{figure*}[!t]
   \centering
   \includegraphics[width=0.40\linewidth]{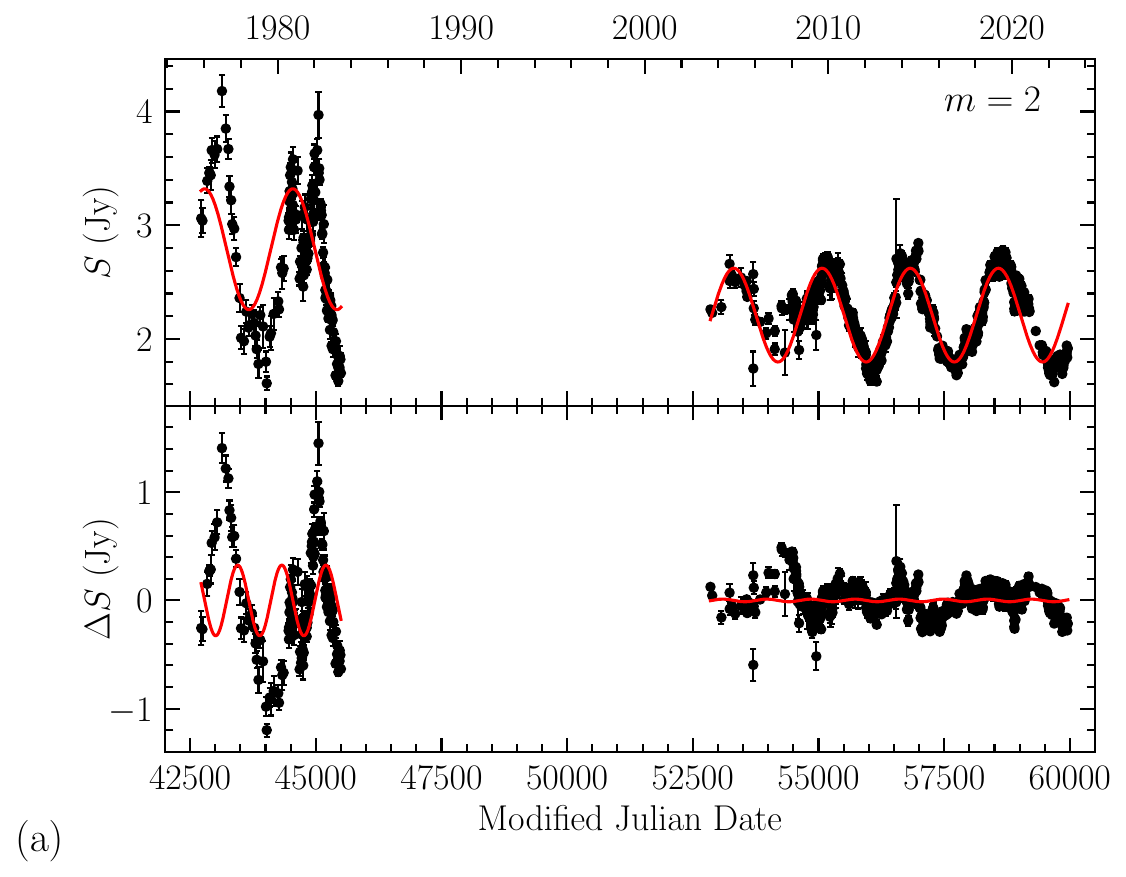}
   \includegraphics[width=0.40\linewidth]{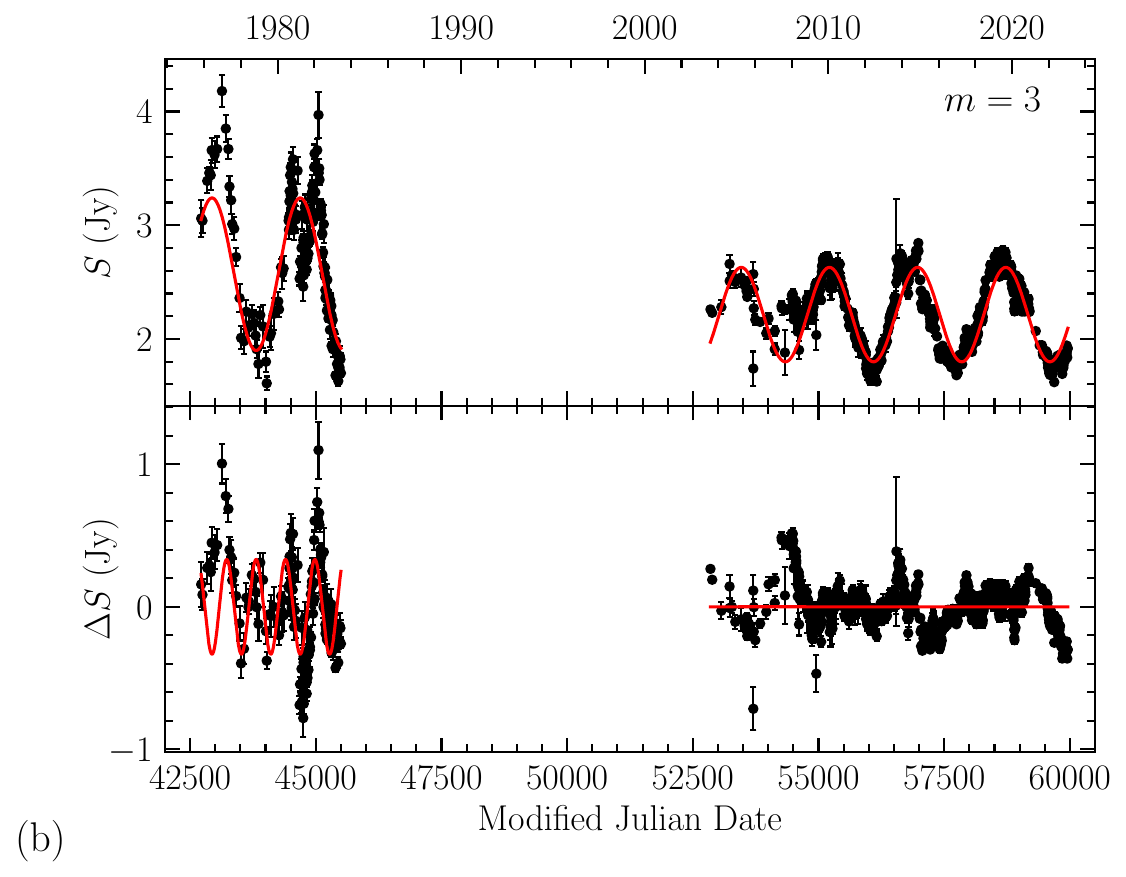}\\
   \includegraphics[width=0.40\linewidth]{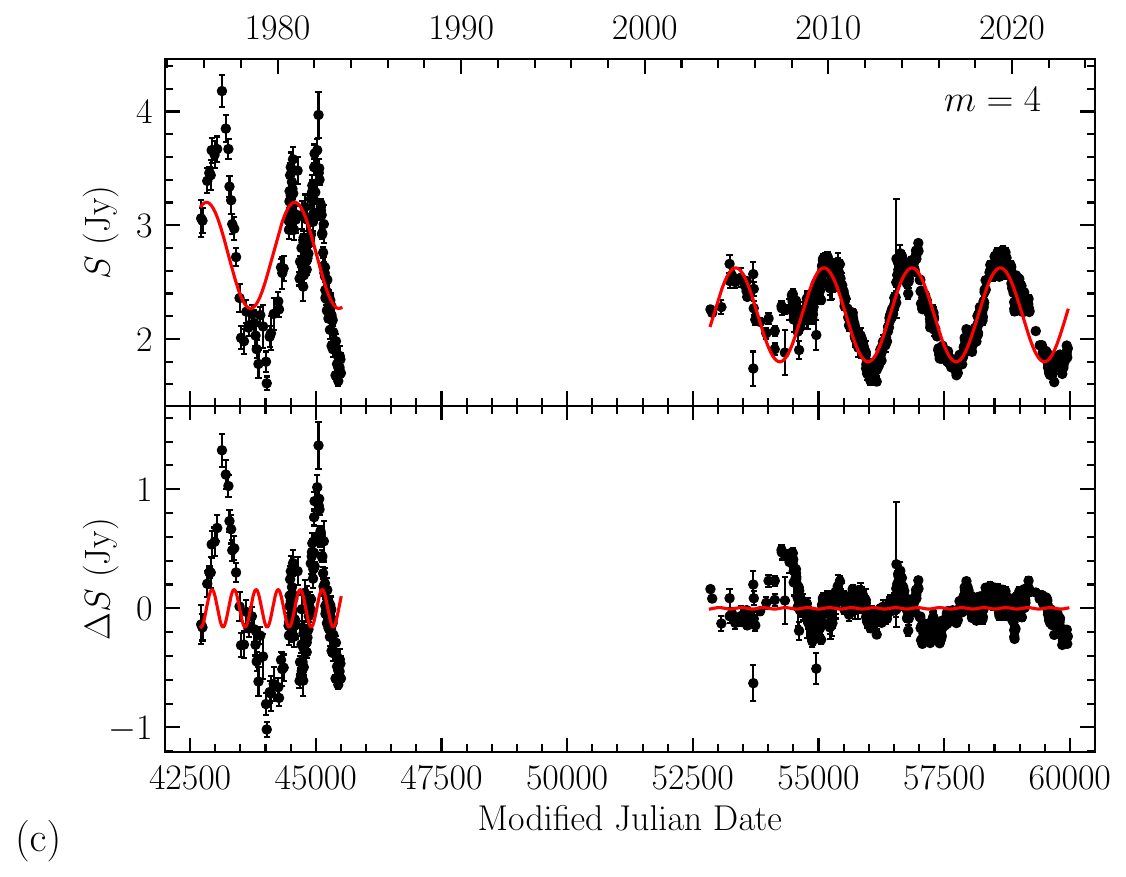}
   \includegraphics[width=0.40\linewidth]{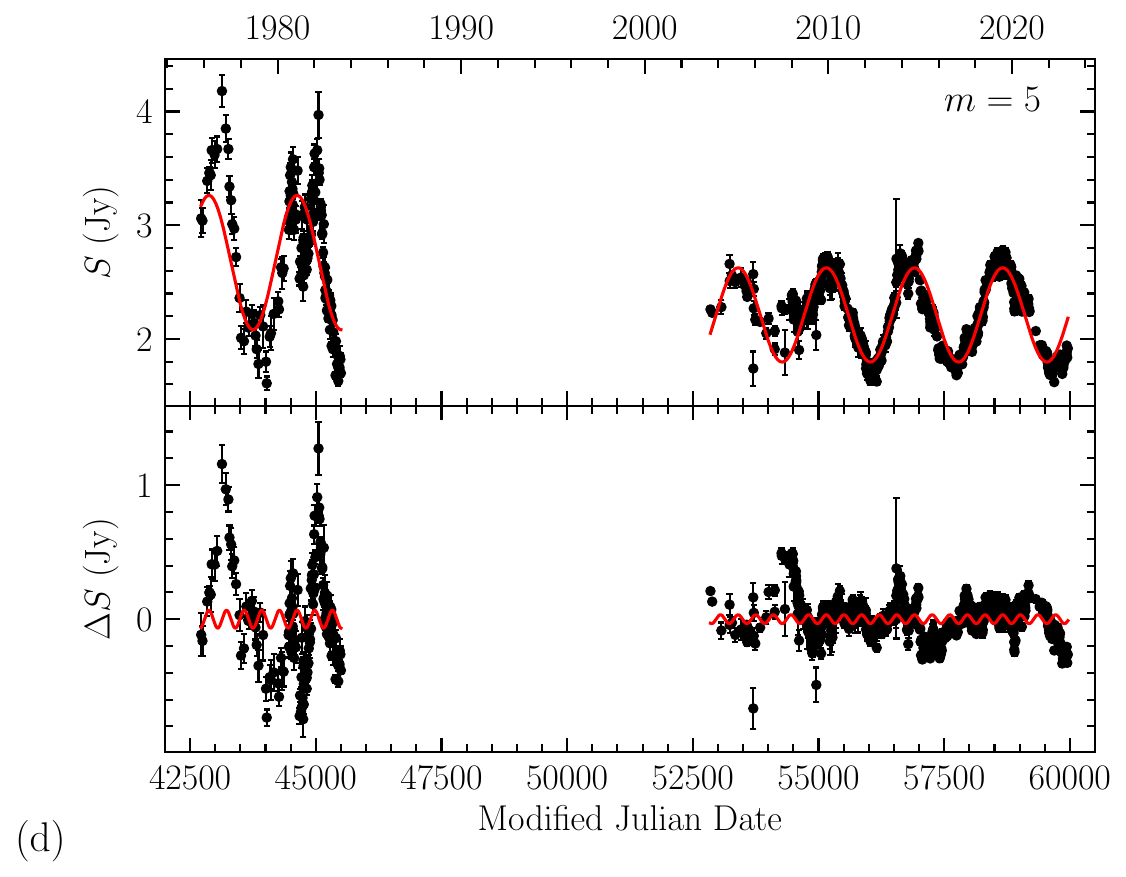}\\
   \includegraphics[width=0.40\linewidth]{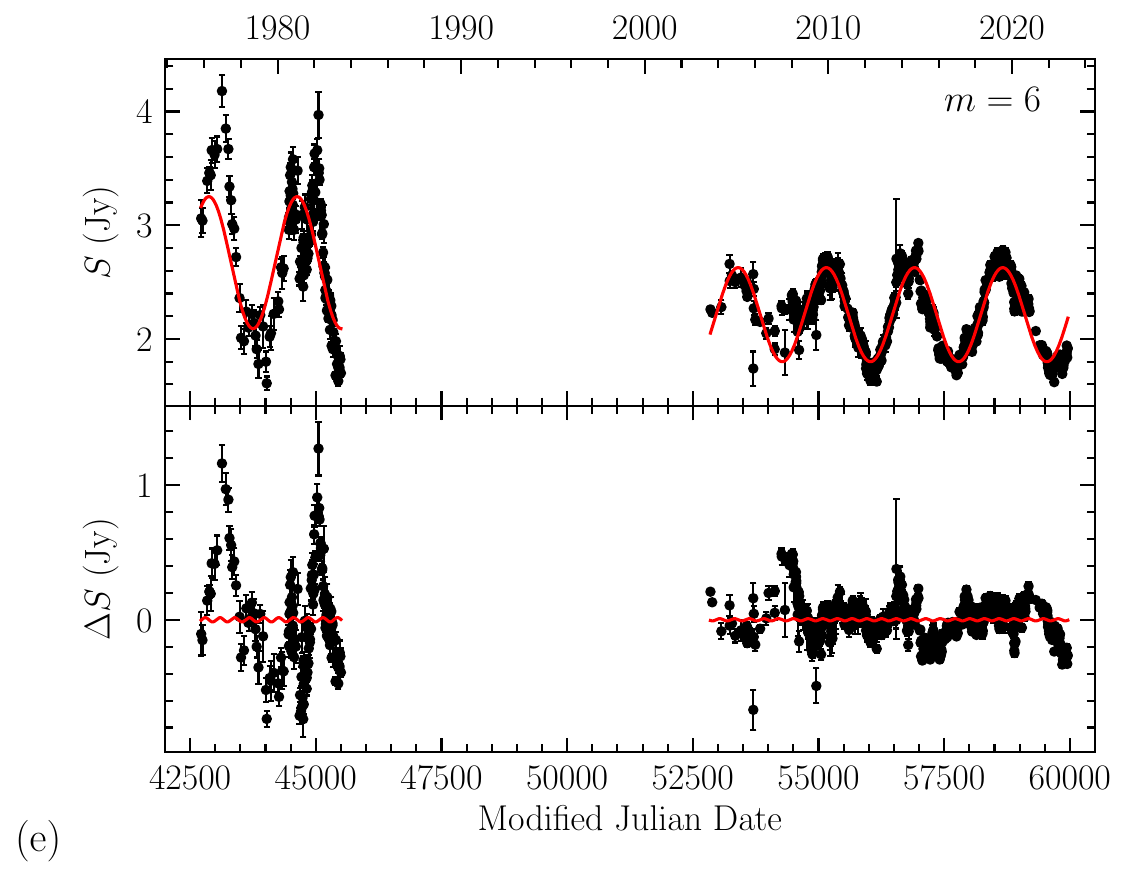}
      \caption{Absence of higher harmonics in the 15~GHz light curve of PKS~2131$-$021. In (a)-(e) the upper panels show the original light curves and the sine wave fits for $m=2$--$6$, and the lower panels show the residual light curves after subtraction of the sine wave fits.}
         \label{plt:harmonics}
\end{figure*}

\begin{table*}[!b]
\centering
\caption{Fit results for the Haystack and OVRO 15~GHz data assuming $t_0=51000$.}
\footnotesize
\begin{tabular}{lcccccc}
\hline \hline
&  no harm. & $m=2$ & $m=3$ & $m=4$ & $m=5$ & $m=6$\\
\hline
$P$ (d)            & $1755.8 \pm 4.9$  & $1776.5 \pm 4.4$  & $1738.1 \pm 1.2$  & $1767.1 \pm 5.0$  & $1755.7 \pm 3.5$  & $1755.8 \pm 4.8$  \\
$\phi_0$ (rad)     & $0.728 \pm 0.064$ & $0.457 \pm 0.056$ & $0.964 \pm 0.016$ & $0.578 \pm 0.064$ & $0.733 \pm 0.046$ & $0.731 \pm 0.063$ \\ 
$A_1$ (Jy)         & $0.563 \pm 0.059$ & $0.528 \pm 0.058$ & $0.666 \pm 0.032$ & $0.478 \pm 0.062$ & $0.566 \pm 0.054$ & $0.562 \pm 0.063$ \\
$A_3$ (Jy)         & \dots             & $0.307 \pm 0.074$ & $0.319 \pm 0.030$ & $<0.239$          & $<0.144$          & $<0.100$          \\
$S_1$ (Jy)         & $2.679 \pm 0.042$ & $2.783 \pm 0.045$ & $2.574 \pm 0.023$ & $2.736 \pm 0.043$ & $2.685 \pm 0.037$ & $2.675 \pm 0.042$ \\
$\sigma_{1}$ (Jy)  & $0.390 \pm 0.033$ & $0.441 \pm 0.030$ & $0.239 \pm 0.018$ & $0.430 \pm 0.030$ & $0.387 \pm 0.030$ & $0.390 \pm 0.033$ \\
$A_2$ (Jy)         & $0.414 \pm 0.007$ & $0.411 \pm 0.006$ & $0.414 \pm 0.007$ & $0.414 \pm 0.006$ & $0.415 \pm 0.006$ & $0.414 \pm 0.007$ \\
$A_4$ (Jy)         & \dots             & $<0.022$          & $<0.010$          & $<0.017$          & $0.032 \pm 0.006$ & $<0.019$          \\
$S_2$ (Jy)         & $2.212 \pm 0.005$ & $2.211 \pm 0.004$ & $2.214 \pm 0.005$ & $2.212 \pm 0.004$ & $2.212 \pm 0.004$ & $2.212 \pm 0.005$ \\
$\sigma_{2}$ (Jy)  & $0.126 \pm 0.004$ & $0.122 \pm 0.003$ & $0.133 \pm 0.004$ & $0.124 \pm 0.004$ & $0.124 \pm 0.003$ & $0.126 \pm 0.004$ \\
$\ln\mathcal{L}_{\rm max}$   & 1359.8  & 1366.9  & 1391.5  & 1361.1 & 1374.6 & 1360.5\\
BIC                & $-2664.3$ & $-2664.7$ & $-2713.9$ & $-2653.1$ & $-2680.1$ & $-2651.9$\\
$\delta$BIC        & 0.0     & $-0.4$    & $-49.6$    & $+11.2$ & $-15.8$ & $+12.4$\\
\hline
\end{tabular}
\label{tab:harmonics}
\end{table*}

To check if any higher harmonic frequencies are present in the data, we fitted the following models to the radio light curve:
$$
S(t)=A_1 \sin(\phi-\phi_0)+A_3 \sin((\phi-\phi_0) m)+S_1$$ 
for MJD $< 45500$, and
$$
S(t)=A_2 \sin(\phi-\phi_0)+A_4 \sin((\phi-\phi_0) m)+S_2
$$
for  MJD $> 52850$,
where $m = 2, 3, 4, 5, 6$ and $\phi=2\pi(t-t_0)/P$, $t_0=51000$. The best-fit parameters were found by maximizing the likelihood function defined as:
\begin{equation}
\ln\mathcal{L}=-\frac{1}{2}\sum_{i}^{t<45500}\left(\frac{(S_i-S(t_i))^2}{\sigma_i^2+\sigma_1^2}+\ln(\sigma_i^2+\sigma_1^2)\right)-\frac{1}{2}\sum_{i}^{t>52800}\left(\frac{(S_i-S(t_i))^2}{\sigma_i^2+\sigma_2^2}+\ln(\sigma_i^2+\sigma_2^2)\right).
\end{equation}

In Fig.~\ref{plt:harmonics} we show the plots of the harmonics fits and in Table~\ref{tab:harmonics} we show the results of the fits.
We used the Bayesian information criterion (BIC) to check if models with harmonic frequencies provide a better fit than a simple sine wave model. The BIC is calculated as follows $\mathrm{BIC}=-2\ln\mathcal{L}_{\rm max}+k\ln n$, where $\mathcal{L}_{\rm max}$ is the maximum value of the likelihood function, $k$ is the number of parameters estimated by the model, and $n$ is the number of data points. For $m=2,\, m=4,\, {\rm and} \,m=6$, $\delta$BIC $\gtrsim 0$, so the simpler model is preferred. The model with $m=3$ results in BIC that is smaller by $\delta$BIC $=-49.6$ than in the simpler model, which formally provides positive evidence that the former model is preferred. However, the improvement comes mostly from Haystack data and the OVRO data do not provide any evidence for the $m=3$ harmonic.  It is safe to assume that this is due to over-fitting of the Haystack data. For $m=5$, the BIC improvement is smaller ($\delta$BIC $=-15.8$), which is likely due to over-fitting of the OVRO data. 

\begin{figure*}
   \centering
   \includegraphics[width=1.0\linewidth]{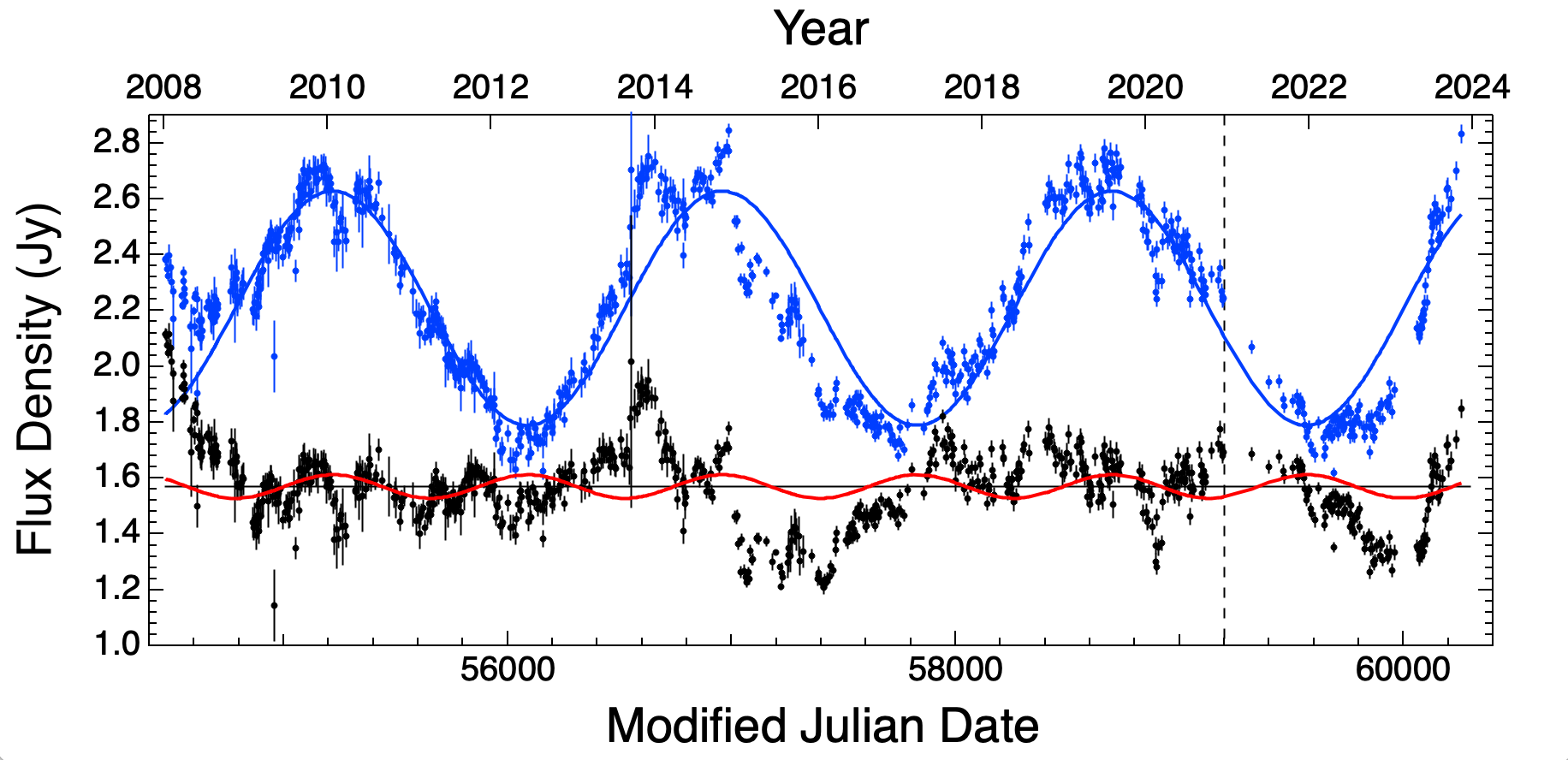}
      \caption{A  first harmonic  of amplitude 0.042 Jy is shown by the red curve.    Sinusoidal first harmonic fluctuations of this amplitude cannot be ruled out by
 our data (see Appendix \ref{sec:brightness}), because of the non-Gaussian nature of the noise, which clearly has long-term correlated fluctuations at this level . Other symbols and curves are the same as those in Fig.~\ref{plt:baselevels}. }
         \label{plt:harmonic}
\end{figure*}

We found the best-fit amplitude of the first harmonic ($m=2$) to be $0.012 \pm 0.006$\,Jy (the 95\% upper limit was 0.022\,Jy). The model with the first harmonic had ${\rm BIC}=-2664.7$, and it was smaller by $\delta({\rm BIC}) = -0.4$ than the BIC of the model without the first harmonic, showing that the former model is slightly, but not significantly, preferred.

 In order to test for a possible first harmonic, we carried out the following exercise. We added an artificial sinusoidal signal with a period of $P/2$,  and an amplitude of 0.012\,Jy to the data. We then fitted sine-wave models (with and without the first harmonic) to the synthetic data. 
The best-fit amplitude of the first harmonic was $0.025 \pm 0.006$\,Jy, and the BIC of the model with the first harmonic was ${\rm BIC} = -2661.7$. The BIC for the model without harmonics was $-2665.0$, so $\delta({\rm BIC}) = +3.3$. When we added an artificial sinusoidal signal with an amplitude of 0.025\,Jy,  the best-fit amplitude of the first harmonic was $0.036 \pm 0.006$\,Jy, and $\delta({\rm BIC}) = -31.3$. In conclusion, we find that an amplitude of the first harmonic of 0.025\,Jy is just consistent (at the $2\sigma$ level) with the data assuming random variations. Given that the variations are clearly not random, as can be seen by the black points in Fig.~\ref{plt:harmonic}, which show long-term correlated fluctuations,  the level of consistency is considerably better than this. So a first harmonic of amplitude 42 mJy, i.e. one tenth of the fundamental, cannot be ruled out  by the data.

\clearpage
\bibliographystyle{aasjournal.bst}
\bibliography{references.bib}

\end{document}